\documentclass{imsart}
\pubyear{2023}
\volume{TBA}
\issue{TBA}
\arxiv{2302.09211}
\firstpage{1}
\lastpage{1}

\usepackage{amsthm}
\usepackage{amsmath}
\usepackage{natbib}
\usepackage[colorlinks,citecolor=blue,urlcolor=blue,filecolor=blue,backref=page]{hyperref}
\usepackage{graphicx}
\usepackage{float}
\usepackage{bbm}

\usepackage{amsfonts,amssymb}

\usepackage{xcolor,colortbl}

\newcommand\bs{\Sigma}
\newcommand\Ybf{\boldsymbol{Y}}

\newcommand\kron{\, \otimes \,}
\newcommand\mname{SWAG }

\definecolor{Gray}{gray}{0.85}

\newcolumntype{a}{>{\columncolor{Gray}}p{.63cm}}
\newcolumntype{x}[1]{>{\centering\arraybackslash\hspace{0pt}}p{#1}}

\graphicspath{ {./Figures/} }

\begin{document}

\begin{frontmatter}
\title{Bayesian Covariance Estimation for Multi-group Matrix-variate Data}
\runtitle{SWAG Covariance Estimation}

\begin{aug}
\author{\fnms{Elizabeth} \snm{Bersson}\thanksref{addr1}\ead[label=e1]{elizabeth.bersson@duke.edu}}
\and
\author{\fnms{Peter D.} \snm{Hoff}\thanksref{addr1}\ead[label=e2]{peter.hoff@duke.edu}}

\runauthor{Bersson and Hoff}

\address[addr1]{Department of Statistical Science, Duke University, Durham, NC;
    \printead{e1} 
    \printead*{e2}
}

\end{aug}

\begin{abstract}

Multi-group covariance estimation for matrix-variate data with small within-group sample sizes
is a key part of many data analysis tasks
in modern applications.
To obtain accurate group-specific covariance estimates, 
shrinkage estimation meth\-ods which
shrink an unstructured, group-specific covariance either
across groups towards a pooled covariance or within each group towards a Kronecker structure
have been developed.
However, in many applications, it is unclear which approach will result in more accurate covariance estimates.
In this article, we present a hierarchical prior distribution which flexibly allows for both types of shrinkage.
The prior linearly combines 
shrinkage across groups towards a shared pooled covariance
and shrinkage within groups towards a group-specific Kronecker covar\-iance.
We illustrate the utility of the proposed prior in speech recognition and an analysis of chemical exposure data.

\end{abstract}


\end{frontmatter}

\section{Introduction}\label{introduction}


Matrix-variate datasets consisting of a sample of $n$ matrices $Y_1,...,Y_n$ each with dimen\-sionality $p_1\times p_2$
are increasingly common in modern applications.
Examples of such datasets include 
repeated measurements of a multivariate response, two-dimensional images, and spatio-temporal observations. 
Often, a matrix-variate dataset may be part\-itioned into several distinct groups  or 
subpopulations
for which group-level inference is of particular interest.
For example, a multi-group dataset may be subdivided by
socio-economic population 
or geographic region.


In analyzing multi-group matrix-variate data of this type, 
describing heterogeneity across groups is often of particular interest.
For instance, in remote-sensing studies, scientists may be interested in understanding variation across classes of land cover
from repeated measurements of spectral information, as in \cite{Johnson2012}.
The information collected at each
site may be represented as a $p_1 \times p_2$ matrix where the rows represent $p_1$ wavelengths and the
columns represent $p_2$ dates when the images were taken.
Accurate multi-group covariance estimation is necessary for such a task.
Moreover,
in many such applications, the population covariance of matrix-variate data may be near separable in that the covariances across the $p_2$ columns are near each other or the covariances across the $p_1$ rows are near each other. 
Incorporation of this structural information in an estimation procedure can improve estimation accuracy.
More generally,
accurate covariance estimation of multi-group matrix-variate data is a 
pertinent task in
many statistical methodologies including classification, principal component analysis, and multivariate regression analysis, among others.
For example, classification of a new observation based on a labeled training dataset
with quadratic discriminant analysis (QDA) requires
group-level estimates of population means and covariance matrices.
As a result, adequate performance of the classification relies on, among other things, accurate group-level covariance estimates.

One approach to analyzing multi-group matrix-variate data is to vectorize each data matrix and utilize methods designed for generic multivariate data separately for each group.
In this way, 
direct covariance estimates which only make use of within-group samples 
may be obtained from matrix-variate data
by vectorizing each
observation
and computing the standard sample covariance matrices. 
While a group's sample covariance may be unbiased, 
the estimate may have large variance
unless the group-specific sample size $n_j$ is appreciably larger than the dimension $p=p_1p_2$.
This is a limiting requirement as
modern datasets often
consist of many features, that is,
often $p\approx n_j$, or even $p\gg n_j$. 
As a result, more accurate covariance estimates may be obtained via indirect or model-based methods which incorporate auxiliary information. Such methods may introduce bias, but can correspond to estimates with lower variance than unbiased methods.



To improve the accuracy of covariance estimates for multi-group data,
researchers may
estimate each group's covariance
with the pooled sample covariance matrix.
This implicitly imposes an assumption of homogeneity of covariances across the populations and greatly reduces the number of unknown parameters to be estimated.
Such an estimator may be biased, but can have lower error than the population-specific sample covariance. 
For example, linear discriminant analysis (LDA), 
which assumes 
homosced\-asticity across groups,
has been shown to outperform QDA when sample sizes are small, even in the presence of heterogeneous population covariances (see, for example, \cite{Marks1974}).
A more robust approach
estimates
each group's covariance as a weighted sum of a pooled estimate and the group-specific sample covariance. 
Such an approach is often referred to as partial-pooling, or, in the Bayesian framework, 
hierarchical modeling.
For a nice introduction to partial pooling and 
an empirical Bayesian implementation, see \cite{Greene1989}.
More work on the topic is found in \cite{Friedman1989,Rayens1991,Brown1999}.
Relatedly, there has been work which assumes pooled elements of common covariance decompos\-itions (e.g. pooled eigenvectors across groups \citep{Flury1987}) and proposals to shrink elements of decompositions to pooled values \citep{Daniels2006,Hoff2009}.


Alternatively, 
as opposed to pooling information across groups,
accuracy may be improved by 
imposing structural assumptions on the 
covariances separately for each group. 
Some common structural assumptions include 
diagonality \citep{Daniels1999}, 
bandability \citep{Wu2009}, and sparsity \citep{Friedman2008}, among others. 
For matrix-variate data, a separable or Kronecker structure covariance assumption \citep{Dawid1981} 
may be more appropriate.
A Kronecker structured covariance  
represents 
each $p\times p$ population covariance 
as the Kronecker product of two smaller covariance matrices of dimension $p_1\times p_1$ and $p_2\times p_2$ which respectively represent the across-row and across-column covariances.
Again, while a separable covariance estimator may be biased,
it may have smaller error than an
unstructured covariance estimator.
In practice, however, 
the population covariance may not be well represented by a Kronecker structure.
To allow for robustness to misspecification, a researcher may proceed in a Bayesian manner and adaptively shrink to the Kronecker structure as in \citet{Hoff2022}. 
Such an estimator can be consistent, but may have stability issues similar to that of an unstructured covariance matrix 
if the population covariance is not well represented by a Kronecker structure.
In this instance, 
more accurate
multi-group covariance estimates
may be obtained by pooling across groups rather than
shrinking within each group to a separable structure.


More generally,
in covariance estimation based on matrix-variate data from multiple populations, it is rarely obvious whether 
shrinking each unstructured covariance separ\-ately towards a Kronecker structure or shrinking all unstructured covariances towards an unstructured pooled covariance 
will result in more accurate estimates.
This is particularly difficult 
in the presence of small within-group sample sizes as
popular
classical statistical tests for both homogeneity of covariances \citep{Box1949} and accuracy of a Kronecker structure assumption \citep{Lu2005}
rely on approximations that require large sample sizes to achieve the desired precision.

To account for this uncertainty, we propose a hierarchical model that adaptively allows for both types of shrinkage.
Specifically, in this article, we
provide a model-based multi-group covariance estimation method 
for matrix-variate data
that improves the overall accuracy of direct covariance estimates.
We propose a hierarchical model for unstructured group-level covariances that adaptively shrinks
each estimate 
either
within-population towards a separable Kronecker structure,
across-populations towards a shared pooled covariance,
or towards a weighted additive combination of the two.
The model features flexibility
in the amount of each type of shrinkage.
Furthermore, the proposed model has a latent-variable representation that results in straightforward Bayesian inference via a Metropolis-Hastings algorithm.
The proposed model provides robustness to mis-specification of structural assumptions and
improved stability if assum\-ptions are wrong while maintaining coherence and interpretability.


The article proceeds as follows. In Section \ref{methodology} we motivate our method and detail the proposed hierarchical model. We describe a Bayesian estimation algorithm in Section \ref{estimation} and demonstrate properties of the proposed method via a simulation study in Section \ref{numerical}.
The flexibility of the proposed method is shown in two examples in Section \ref{examples}. 
In the first example, we demonstrate the usefulness of inference under the proposed model in speech recognition. In the second example, we perform inference on a chemical exposure data set where understanding heterogeneity across socio-demographic groups is of key interest. 
We conclude with a discussion in Section \ref{discussion}.

\section{Methodology}\label{methodology}

In this section we introduce the
\textit{Shrinkage Within and Across Groups} (SWAG) covariance model, a hierarchical model developed for simultaneous covariance estimation
for multi-group, matrix-variate data.
We are particularly motivated by improving the overall accuracy of group-specific estimates of population covariances
when the true covariance structures are unknown and group-specific sample sizes
are small
relative to the number of features. 
The proposed model adaptively allows for flexible shrinkage either across groups, within a group to a Kronecker structure, or an additive combination of the two.
The \mname model is constructed from semi-conjugate priors to allow for straightforward Bayesian estimation and interpretable parameters.
The section proceeds by introducing motivation in Sections \ref{shrinkpoolsec} and \ref{shrinkkronsec}, presenting the proposed hierarchical covariance model in Section \ref{swagsec}, and elaborating on parameter interpretation in Section \ref{swagintsec}.

\subsection{Partial-pooling shrinkage for multi-group data}\label{shrinkpoolsec}

As detailed in the Introduction, 
a common method used to improve 
a population's covariance estimate
is
linear shrinkage from the population's sample covariance matrix towards some covariance term which can be estimated with greater precision.
One such method for multi-group multivariate data is partial pooling, as detailed in \citet[][GR]{Greene1989}. 
In particular, 
for population $j\in\{1,...,J\}$, let $y_{1,j},...,y_{n_j,j}$ be an i.i.d. random sample of $p$-dimensional vectors from a mean-zero normal population with unknown non-singular covariance matrix $\bs_j\in\mathcal{S}^+_p$,
\begin{equation*}
      y_{1,j},...,y_{n_j,j}\sim N_{p}\left(0,\Sigma_j\right),\quad\text{independently for }i=1,...,n_j,\;j =1,...,J.
\end{equation*}
GR use mutually independent inverse-Wishart priors for each population covariance,  \\
$\bs_j^{-1}\sim \text{Wishart}_p\big(\Psi_0^{-1}/(\nu-p-1),\nu\big),$ for $j\in\{1,...,J\}$, parameterized such that $E[\bs_j|\Psi_0,\nu]=\Psi_0$. 
The Bayes estimator for the covariance of population $j$ under squared error loss partially-pools each group's sample covariance,
\begin{equation}\label{bayespp}
    \hat{\bs}_j := E\left[\bs_j|\boldsymbol{y},\Psi_0,\nu\right] =
   w_1 S_j + (1-w_1){\Psi}_0,
\end{equation}
where $w_1= n_j/(n_j+{\nu}-p-1)$ and
$S_j=\sum_{i=1}^{n_j}y_{i,j}y_{i,j}^T/n_j$ is the sample covariance matrix for population $j$.
GR use plug-in estimates for the pooled covariance and degrees of freedom $\{\Psi_0,\nu\}$ which are obtained in an empirical Bayesian manner.

This partially-pooled estimator linearly shrinks a population's sample covariance matrix towards a pooled covariance by a weight $w_1$
that depends on the degrees of freedom and the group-specific sample size $n_j$.
In this way, the degrees of freedom parameter determines the amount of shrinkage towards the pooled covariance.
In partic\-ular, if the degrees of freedom $\nu$ is large relative to the sample size $n_j$, 
the covariance estimate is strongly shrunk towards the pooled value. 
In populations where the group-specific sample size is large, or if the degrees of freedom estimate is comparatively small, 
more weight is placed on
the sample covariance matrix.

\subsection{Kronecker shrinkage for matrix-variate data}\label{shrinkkronsec}

For a matrix-variate population,
the accuracy of a covariance estimate may be improved via linear shrinkage towards a population-specific Kronecker structure.
Let $Y_{1},...,Y_{n}$ be an i.i.d. sample of random matrices, each of dimension $p_1\times p_2$ , from a mean-zero normal population with non-singular covariance matrix $\bs\in\mathcal{S}^+_p$ where $p=p_1p_2$,
\begin{equation*}
    Y_{1},...,Y_{n}\sim N_{p_1\times p_2}\left(0,\bs\right),\quad\text{independently for }i=1,...,n.
\end{equation*}
Even if $p_1$ and $p_2$ are each relatively small, 
obtaining a statistically stable estimate of
the unstructured $p$-dimensional covariance 
may require a prohibitively large sample size.
As a result, 
shrinkage towards 
a parsimonious Kronecker structured covariance $C\kron R$ may be used,
where 
``$\kron$'' is the Kronecker product,
$R\in\mathcal{S}^+_{p_1}$ is a ``row'' covariance matrix,
and $C\in\mathcal{S}^+_{p_2}$ is a ``column'' covariance matrix.
A linear shrinkage estimator
that shrinks a population's sample covariance towards a population-specific Kronecker separable covariance may be obtained from the following prior,
\begin{equation*}
    \bs^{-1} \sim  \text{Wishart}_{p}\left(\left(C\kron R\right)^{-1}/(\gamma-p-1),\gamma\right),
\end{equation*}
parameterized so that $E[\bs_j|C,R,\gamma] = C\kron R$.
Here, the Bayes estimator of the covariance $\bs$ under squared error loss is 
\begin{equation}\label{kronest}
    \hat{\bs} :=  E\left[\bs|\boldsymbol{Y}, C, R, \gamma\right] =  w_2S+ (1-w_2)\left({C}\kron{R}\right),
\end{equation}
where $w_2=n/(n+{\gamma}-p-1)$. 
An empirical Bayesian estimation approach
based on shrinkage towards a Kronecker structure
is presented in \cite{Hoff2022}.
In context to that article, here, we will take a
fully Bayesian approach.

The estimator given in Equation \ref{kronest}
linearly shrinks the unstructured sample covariance towards a Kronecker structured covariance by the weight $w_2$ that depends on the sample size and the estimated degrees of freedom.
As with the partially-pooled estimator, this estimator is strongly shrunk towards the Kronecker structure when the degrees of freedom is large relative to sample size.
If the degrees of freedom is small, or the sample size is large, more weight is placed on the sample covariance.

\subsection{Flexible shrinkage for multi-group matrix-variate data}\label{swagsec}

For each group $j=1,...,J$, let $Y_{j,1},...,Y_{j,n_j}$ be an i.i.d.
sample of random matrices, each of dimension $p_1\times p_2$, from a  mean-zero normal population with non-singular covariance $\bs_j\in\mathcal{S}^+_p$, that is, 
\begin{equation}\label{samplingdist}
    Y_{j,1},...,Y_{j,n_j}\sim N_{p_1\times p_2}\left(0,\bs_j\right),\quad\text{independently for }j =1,...,J.
\end{equation}
As it is often unclear 
which of the approaches presented
is most appropriate in the presence of multi-group matrix-variate data of this type, 
we propose an approach that combines the two methods of covariance shrinkage discussed.
In particular,
we propose
the
\textit{Shrinkage Within and Across Groups} (SWAG)
hierarchical prior distribution
which
linearly combines an estimate shrunk towards a pooled covariance $\Psi_0$ and 
an estimate shrunk towards a group-specific Kronecker structure $\left(C_j\kron R_j\right)$
by a weight $\lambda\in(0,1)$ that is estimated from the data.

\begin{figure}[ht]
\centering
\includegraphics[width=12cm,keepaspectratio]{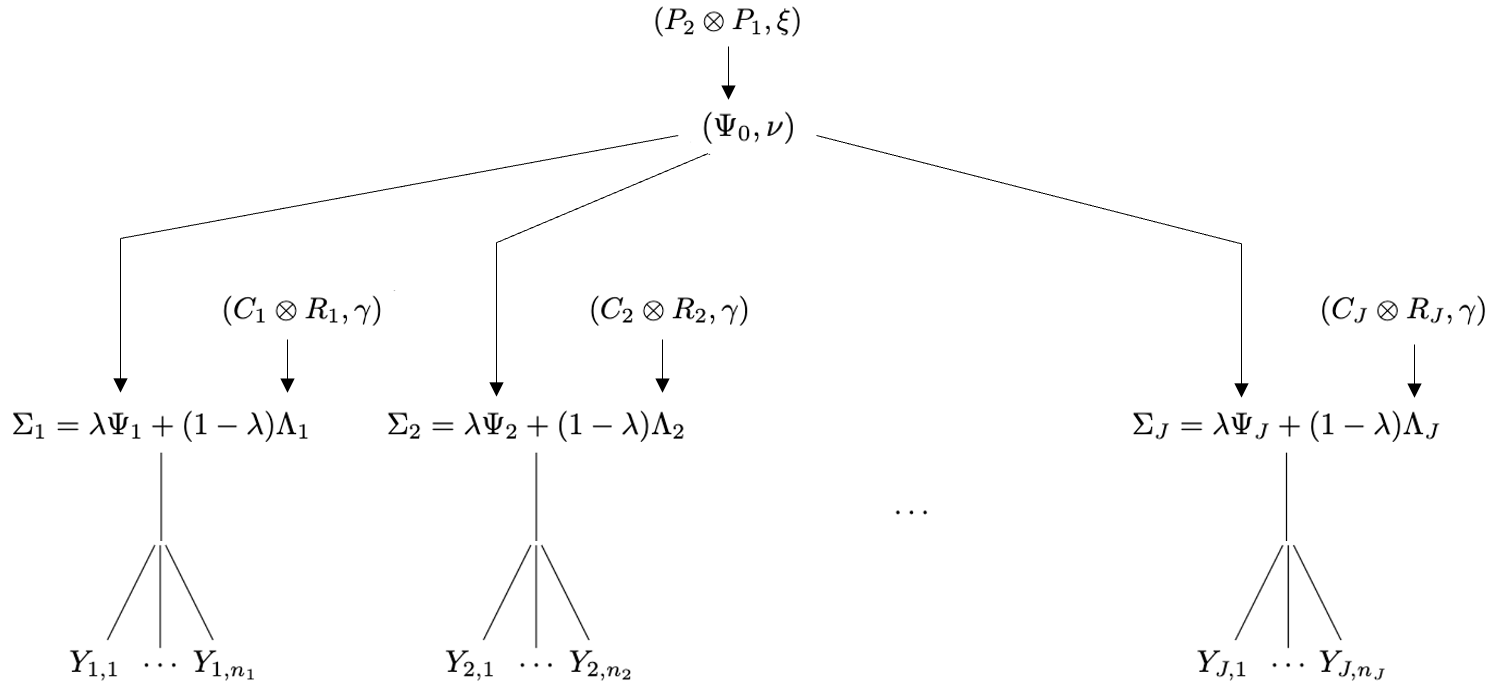}
\caption{
A graphical representation of the SWAG hierarchical model.
The unstructured covariance terms $\Psi_1,...,\Psi_J$ are shrunk across groups towards a shared covariance $\Psi_0$. 
The covariance terms $\Psi_0,\Lambda_1,....\Lambda_J$ are each individually shrunk towards a Kronecker covariance.  
}\label{hierarchy}
\end{figure}

Specifically, the \mname prior utilizes an over-parameterized representation of each group's covariance. That is, for population $ j\in\{1,...,J\}$,
\begin{align}
    \Sigma_j ={}& \lambda\Psi_j + (1-\lambda)\Lambda_j,\label{overparam}
\end{align}
where each $\Psi_j$ is shrunk towards a common covariance, and each $\Lambda_j$ is shrunk towards a group specific Kronecker covariance.
Each covariance $\Psi_j$ is shrunk across groups
towards a common covariance matrix using the prior distribution
\begin{align}
    \Psi_j^{-1}\sim{}&  \text{Wishart}_{p}\left(\Psi_0^{-1}/(\nu-p-1),\nu\right),\quad\text{independently for }j=1,...,J \label{shrinkpool},
\end{align}
parameterized such that $E[\Psi_j|\Psi_0,\nu]=\Psi_0$.
When $\Psi_0$ is estimated from data across all groups, this term is interpreted as a pooled covariance matrix. 
As we are interested in obtaining a covariance matrix estimate based on matrix-variate data, each $\Lambda_j$ term is shrunk towards a group-specific Kronecker structured covariance, 
\begin{align}
    \Lambda_j^{-1}\sim{}&  \text{Wishart}_{p}\left(\left(C_j\kron R_j\right)^{-1}/(\gamma-p-1),\gamma\right),\quad\text{independently for }j=1,...,J,\label{shrinkkron}
\end{align}
where $E[\Lambda_j|R_j,C_j,\gamma] = \left(C_j\kron R_j\right)$.
Here, as before, $R_j$ is a $p_1\times p_1$ row covariance matrix from population $j$ and 
$C_j$ is the corresponding $p_2\times p_2$ column covariance matrix. 
Furthermore, 
to more clearly separate these two notions of shrinkage (within population or across populations), 
a Wishart prior on the across-population covariance $\Psi_0$ 
allows for flexible shrinkage of this pooled term towards an across-group Kronecker covariance:
\begin{equation}
    \Psi_0 \sim \text{Wishart}_p\left(\left(P_2\kron P_1\right)/\xi,\xi\right),\label{shrinkpoolkron}
\end{equation}
parameterized such that $E\left[\Psi_0|P_1,P_2,\xi\right] = \left(P_2\kron P_1\right)$ where $P_1\in\mathcal{S}^+_{p_1}$ and $P_2\in\mathcal{S}^+_{p_2}$. 
In this way, the weight $\lambda$ is interpreted
as partially controlling the amount of 
shrinkage towards
homogeneity versus towards heterogeneity of covariances across groups. 
A visualization of the proposed \mname hierarchy is given in Figure \ref{hierarchy}. In summary, the \mname model combines a standard hierarchical model on the across-group shrunk $\Psi_j$ covariances with Bayesian shrinkage towards a separable structure on the within-group shrunk $\Lambda_j$ covariances and the pooled covariance $\Psi_0$. 

We note that 
the \mname model is primarily motivated by the need to obtain more accurate group-specific covariance estimates,  so,
while there is redundancy in this param\-eterization,
the group-specific covariances $\bs_1,...,\bs_J$ are identifiable.  As a result,  this over-parameterization will not affect inference on estimation of group-level covariances, estimation of mean effects, imputation of missing data, or response prediction. 

\subsection{Interpretation of Parameters}\label{swagintsec}

In this section, we highlight properties of the proposed \mname covariance priors under the normal sampling model given in Equation \ref{samplingdist}.
A priori, regardless of the specified sampling model, the marginal expectation under the \mname prior is
a weighted sum
of a pooled covariance and the heterogeneous Kronecker separable covariance:
\begin{equation}\label{priormargex}
    E\left[\bs_j|\Psi_0,R_j,C_j,\lambda\right] = \lambda\Psi_0 + (1-\lambda)\left(C_j\kron R_j\right),
\end{equation}
for $j\in\{1,...,J\}$, where the weight $\lambda$ quantifies the prior weight on 
each structure.
As a result, the \mname prior presents a flexible approach to combine shrinkage across groups towards a pooled value and shrinkage within groups towards a separable structure.

The role of the weight $\lambda$ is
further elucidated at the extremes of its sample space.
Given $\lambda=1$, the \mname model reduces to a Bayesian analogue of the partially-pooled estimators given in \cite{Greene1989}, as given in Equation \ref{bayespp}.
At the weight's opposite extreme, when $\lambda=0$, 
the \mname model 
is equivalent to the Kronecker structure shrinkage model presented 
in Section \ref{shrinkkronsec} applied separately to each group.
To alleviate some of the potential ambiguity in interpretation of $\lambda$, we further allow $\Psi_0$ to be flexibly shrunk to a Kronecker structure. 
In this way, 
$\lambda=1$ represents hierarchical shrinkage across populations towards a pooled covariance and $\lambda=0$ represents within-population shrinkage.

Moreover, under the \mname prior, the prior marginal expected value of a population's covariance is a weighted sum of a pooled covariance and a population-specific separable covariance (Equation \ref{priormargex}).
Under normality, as the degrees of freedom $\nu$ and $\gamma$ increase, the marginal sampling model of a random matrix $Y_{i,j}$ converges to a normal distribution with this prior expectation as the covariance.
That is, when the degrees of freedom parameters $\nu$ and $\gamma$ are large, 
the unstructured covariances $\Psi_j$ and $\Lambda_j$
are each
 strongly shrunk towards their respective prior expected value, and, therefore, 
the population covariances are approximately represented by the weighted sum of the pooled covariance and population-specific Kronecker structure.

\section{Parameter Estimation}\label{estimation}

\subsection{Latent Variable Representation}

The \mname model has a latent variable representation 
that
allows for straightforward Bayesian inference. 
Specifically, consider the following representation of the proposed \mname model:
\begin{align}\label{latentrep}
     vec(Y_{i,j})={}& \lambda^{1/2}U_{i,j} + (1-\lambda)^{1/2}E_{i,j},\quad \text{for }i=1,...,n_j\\
     U_{i,j}\sim{}& N_{p}\left(0,\Psi_j\right)\nonumber\\
     E_{i,j}\sim{}& N_{p}\left(0,\Lambda_j\right),\nonumber
\end{align}
independently for each population $j\in\{1,...,J\}$
where the priors on each $\Psi_j$ and $\Lambda_j$ are as given in Equations \ref{shrinkpool} and \ref{shrinkkron}.
That is, each matrix $Y_{i,j}$ is represented as a weighted sum of 
one factor ($U_{i,j}$) 
which partially pools covariances across populations
and another factor ($E_{i,j}$) which flexibly shrinks the population covariance towards a group-specific Kronecker structure.
Marginal with respect to the factors, this latent variable representation is equivalent to the sampling model proposed in the \mname model,
\begin{equation*}
    vec(Y_{i,j})|\lambda,\Psi_j,\Lambda_j \sim N_{p}\left(0, \lambda\Psi_j +(1-\lambda)\Lambda_j\right).
\end{equation*}
Conditioning on the factor $U_{i,j}$ 
results in
closed form full conditionals of the covariance parameters of interest, as detailed in subsequent subsections.

\subsection{Posterior Approximation}\label{fullcond}

In this section, we detail a Metropolis-Hastings 
algorithm for parameter estimation based on the latent variable representation of the \mname model.
Label $\boldsymbol Y =\{Y_{i,j}: i\in \{1,...,n_j\},j\in\{1,...,J\} \}$,
$\boldsymbol U =\left\{U_{i,j}: i\in \{1,...,n_j\},j\in\{1,...,J\} \right\}$,  $\boldsymbol \Psi = \{\Psi_1,...,\Psi_J\}$, $\boldsymbol \Lambda = \{\Lambda_1,...,\Lambda_J\}$, $\boldsymbol{R}=\{R_1,...,R_J\}$, and $\boldsymbol{C}=\{C_1,...,C_J\}$.
Then, Bayesian inference is based on the joint posterior distribution, with density
\begin{equation*}
    p(\lambda,\boldsymbol{\Psi},\boldsymbol{\Lambda},\Psi_0,\nu,\boldsymbol{R},\boldsymbol{C},\gamma,P_1,P_2,\xi|\boldsymbol{Y}),
\end{equation*}
and a Monte Carlo approximation to this posterior distribution is available via a Metropolis\-Hastings algorithm.
Based on the latent variable representation presented in Equations \ref{latentrep}, 
nearly all of the parameters in the \mname model maintain semi-conjugacy leading to a straightforward Metropolis-Hastings algorithm
which constructs a Markov chain in
\begin{equation*}
    \boldsymbol{\theta} = \{\lambda,\boldsymbol{\Psi},\boldsymbol{\Lambda},\boldsymbol{U},\Psi_0,\nu ,\boldsymbol{R},\boldsymbol{C},\gamma,P_1,P_2,\xi\}.
\end{equation*}
Such Bayesian inference provides estimates and uncertainty quantification 
for arbitrary functions of the parameters.
While we focus on the mean-zero case, the algorithm presented may be trivially extended to include estimation of mean parameters.

For a full Bayesian analysis, priors must be specified for all unknown parameters. 
For simplicity, a straightforward $beta(\alpha,\beta)$ prior may be used to describe prior expectations of behavior of $\lambda$. 
On the degrees of freedom parameters $\nu,\gamma,$ and $\xi$, 
negative binomial distributions with the appropriate support may be used, that is,
$\text{NegBin}_{[p+2,\infty)}(r_0,p_0)$, 
parameterized by size $r_0$, success probability $p_0$, and lower bound $p+2$.
Semi-conjugate priors on the remaining covariance parameters are proposed to facilitate computation,
\begin{align*}
    R_1,...,R_J\sim{}& \text{Wishart}_{p_1}\left(R_0/\eta_1,\eta_1\right)\\
    C_1,...,C_J \sim{}& \text{Wishart}_{p_2}\left(C_0/\eta_2,\eta_2\right)\\
    P_1^{-1}\sim{}&  \text{Wishart}_{p_1}\left(P_{01}^{-1}/(\eta_3-p_1-1),\eta_3\right)\\
    P_2^{-1}\sim{}&  \text{Wishart}_{p_2}\left(P_{02}^{-1}/(\eta_4-p_1-1),\eta_4\right).
\end{align*}
A discussion of hyperparameter specification is provided in Section \ref{priorspec}.

A Metropolis-Hastings sampler proceeds by iteratively
generating new sets of model parameters based on their full conditional distributions.
When iterated until convergence, this procedure will
generate a Markov chain that approximates the joint posterior distribution $p(\boldsymbol{\theta}|\boldsymbol{Y})$.
The sampling steps are now detailed.

\subsubsection{Sampling of population model parameters}

The full conditionals of sampling model covariances and factors, as well as the key shrinkage controlling parameters $\lambda, \nu,\gamma$ are discussed in this subsection.
To reduce dependence along the Markov chain, we propose to sample the weight $\lambda$ and the latent
factor $\boldsymbol{U}$ from their joint full
conditional distribution, $p(\lambda,\boldsymbol{U}|\boldsymbol{Y},\boldsymbol{\theta}_{-\boldsymbol{U},-\lambda})$, where, to streamline notation, we let $ \boldsymbol{\theta}_{-(\cdot)}$ denote the set containing all variables in $\boldsymbol{\theta}$ except for $(\cdot)$. 
A sample from $p(\lambda|\boldsymbol{Y},\boldsymbol{\theta}_{-\boldsymbol{U},-\lambda})$ must first be obtained where
\begin{align*}
        p(\lambda|\boldsymbol{Y},\boldsymbol{\theta}_{-\boldsymbol{U},-\lambda})\propto \prod_{j=1}^J[ 
        \left|\lambda\Psi_j + (1-\lambda)\Lambda_j\right|^{-n_j/2}
        e^{tr(- Y_j\left(\lambda\Psi_j + (1-\lambda)\Lambda_j\right)^{-1}Y_j^T )}]
        \lambda^{\alpha-1}(1-\lambda)^{\beta-1}.
\end{align*}
As the full conditional distribution of $\lambda$ is not available in closed form, a sample of $\lambda$ may be obtained from a Metropolis step that proceeds by first obtaining a proposed value  $\lambda^{*}$ from a reflecting random walk around the previous value of $\lambda$ in the Markov chain \citep{Hoff2009a}. 
That is, an initial value is sampled from $\lambda^* \sim \text{Uniform}(\lambda-\delta_\lambda,\lambda+\delta_\lambda)$, and
it is reflected across the approporiate bound to retain the correct support, $\lambda\in(0,1)$:
\[
\lambda^{*} =
\begin{cases}
\lambda^{*}&\text{ if  } \lambda^{*}\in(0,1),\\
|\lambda^{*}|&\text{ if  }\lambda^{*}\leq0,\\
2-\lambda^{*}&\text{ if }\lambda^{*}\geq 1.
\end{cases}
\]
Then, the proposal $\lambda^*$ is accepted as an updated value for $\lambda$ with probability $r = p(\lambda^*|\boldsymbol{Y},\boldsymbol{\theta}_{-\boldsymbol{U},-\lambda})/ 
p(\lambda|\boldsymbol{Y},\boldsymbol{\theta}_{-\boldsymbol{U},-\lambda}).$
The full conditional of each latent factor $U_j$ for populat\-ion $j$ is independ\-ently $N_{n_j\times p}\left(M_j,S_j\kron I_{n_j}\right)$ where
 \begin{align*}
            S_j={}& \left(\Psi_j^{-1}+\frac{\lambda}{1-\lambda}\Lambda_{j}^{-1}\right)^{-1}\\
            M_j ={}& \frac{\lambda^{1/2}}{1-\lambda}Y_j\Lambda_j^{-1}S_j.
\end{align*}
Again, to reduce dependence along the Markov chain, we propose to sample the degrees of freedom
$\nu$ and covariances $\boldsymbol{\Psi}$ as well as $\gamma$ and $\boldsymbol{\Lambda}$ from their joint full conditional distribution.
 In particular, the joint full conditional of $(\nu,\boldsymbol{\Psi})$ is 
 $p(\nu,\boldsymbol{\Psi}|\boldsymbol{Y},\boldsymbol{\theta}_{-\boldsymbol{\Psi},-\nu}) = \\ p\left(\boldsymbol{\Psi} |\boldsymbol{Y},\boldsymbol{\theta}_{-\boldsymbol{\Psi}}\right) \times $ $ p\left(\nu |\boldsymbol{Y},\boldsymbol{\theta}_{-\boldsymbol{\Psi},-\nu}\right)$
 where $\nu$ may be sampled from a reflecting random walk Metropolis step. 
In this case, a proposal $\nu^*$ may be obtained from a reflecting random walk based on the previous iteration's value of $\nu$, that is,
sample an initial value from $\nu^*\sim \text{Uniform}(\nu-\delta_\nu,\nu+\delta_\nu)$ and utilize the following reassignment schema to ensure the sample has the correct support:
\[
\nu^* = \begin{cases}
\nu^* &\text{if }\nu^*\geq p+2\\
(p+2)+(p+2-\nu^*) &\text{if }\nu^*< p+2.
\end{cases}
\]
The proposal $\nu^*$ is accepted as an updated value for $\nu$ with probability
\[
r = \prod_{j=1}^J\frac{p(U_j|\Psi_0,\nu=\nu^*)p(\nu=\nu^*|r_0,p_0)}{p(U_j|\Psi_0,\nu=\nu)p(\nu=\nu|r_0,p_0)}
\]
where $U_j|\Psi_0,\nu\sim\text{T}_{n_j\times p}\left(\nu-p+1,0,\Psi_0(\nu-p-1)\kron I_{n_j} \right)$.
Then, sample each $\Psi_j$ from its full conditional distribution,
\[
\Psi_j^{-1}|\boldsymbol{Y},\boldsymbol{\theta}_{-\Psi_j}\sim  \text{Wishart}_p\big(\left(U_j^TU_j+(\nu-p-1)\Psi_0\right)^{-1},\nu+n_j\big)
\]
 for each $j\in\{1,...,J\}$.
Samples from
the joint full conditional distribution of $(\gamma,\boldsymbol{\Lambda})$, 
$p(\boldsymbol{\Lambda}|\Ybf,\boldsymbol{\theta}_{-\boldsymbol{\Lambda}}) $ $\times $ $
 p\left(\gamma |\Ybf,\boldsymbol{\theta}_{-\boldsymbol{\Lambda},-\gamma}\right)$
are obtained similarly.
A proposal sample $\gamma^*$ is obtained from a reflecting random walk based on an initial value drawn from $\text{Uniform}(\gamma-\delta_\gamma,\gamma+\delta_\gamma)$
and accepted with probability
\[
r = \prod_{j=1}^J\frac{p\left(Y_j|\lambda,U_j,R_j,C_j,\gamma=\gamma^*\right)p(\gamma=\gamma^*|r_0,p_0)}{p\left(Y_j|\lambda,U_j,R_j,C_j,\gamma=\gamma\right)p(\gamma=\gamma|r_0,p_0)}
\]
where $Y_j|\lambda,R_j,C_j \sim T_{n_j\times p}\left(\gamma-p+1,\lambda^{1/2}U_j,(1-\lambda)\left(C_j\kron R_j\right)(\gamma-p-1)\kron I_{n_j}\right)$, 
and the full conditional of each $\Lambda_j^{-1}$ is independently 
$\text{Wishart}_p\big(\big(\tilde{Y}_j^T\tilde{Y}_j/(1-\lambda) + \left(C_j\kron R_j\right)(\gamma-p-1) \big)^{-1},\gamma+n_j \big)$
where $\tilde{Y}_j = \left(Y_j-\lambda^{1/2}U_j\right)$.

In addition to facilitating computation, these full conditionals contribute to the interpretation of parameters. In particular, for population $j\in\{1,...,J\}$, the full condit\-ional means of $\Psi_j$ and $\Lambda_j$  resemble shrinkage estimators towards a pooled covariance and a Kronecker structured covariance, respectively.
Specifically,
the full conditional density of
the across-group shrunk covariance
$\Psi_j$,
given all other model parameters and the observed data matrices $\Ybf$, is inverse-Wishart with mean
\begin{align}
    E\left[\Psi_j|\boldsymbol{Y},\boldsymbol{\theta}_{-\Psi_j}\right] ={}&w_1U_j^TU_j/n_j+ (1-w_1)\Psi_0,\label{pmPsi}
\end{align}
where
$w_1 = n_j/(n_j+\nu-p-1)$.
That is, the prior on $\Psi_j$ shrinks the sample covariance of the latent factor $U$ towards the pooled covariance by a shrinkage factor determined by the degrees of freedom $\nu$ and the within-group sample size $n_j$.
Similarly, the full conditional density of the within-group shrunk $\Lambda_j$
is inverse-Wishart with mean
\begin{align}
    E\left[\Lambda_j|\Ybf,\boldsymbol{\theta}_{-\Lambda_j}\right] ={}& w_2 \tilde{Y}_j ^T\tilde{Y}_j/n_j+ (1-w_2)\left(C_j\kron R_j\right),\label{pmLambda}
\end{align}
where $ \tilde{Y}_j = \left(Y_j-\lambda^{1/2}U_j\right)/\sqrt{1-\lambda}$ and
$w_2 = n_j/(n_j+\gamma-p-1)$.
In this case, the prior on $\Lambda_j$ shrinks the sample  covariance of the data residual towards a separable covariance by a weight which depends on the degrees of freedom $\gamma$ and $n_j$.

\subsubsection{Full conditionals of Kronecker shrinkage parameters}

The unstructured covariances $\boldsymbol{\Lambda}$ are each shrunk towards a population-specific Kronecker structured covariance.
The derivations of the full conditionals of these Kronecker covar\-iances make use of a few Kronecker product properties, namely,
$tr( B^TA_1BA_2^T) = vec(B)^T(A_2\kron A_1)vec(B)$ and $|A_2\kron A_1| = |A_1|^{p_2}|A_2|^{p_1}$ for $A_1$ of dimension $p_1\times p_1$ and $A_2$ of dimension $p_2\times p_2$.  
Then, it is straightforward to derive the full conditional of each population $j$'s row covariance,
\[
R_j\sim \text{Wishart}_{p_1}\big(\left((\gamma-p-1)\sum_{k=1}^pL_kC_jL_k^T + R_0^{-1}\eta_1\right)^{-1},\eta_1+\gamma p_2\big)
\]
for $L_k=vec^{-1}(l_k)$ from $\Lambda_j^{-1}=\tilde{L}\tilde{L}^T=\sum_{k=1}^pl_kl_k^T$
and column covariance
\[
C_j\sim \text{Wishart}_{p_2}\big(\left((\gamma-p-1)\sum_{k=1}^pL_k^TR_jL_k + C_0^{-1}\eta_2\right)^{-1},\eta_2+\gamma p_1\big)
\]
for 
$L_k=vec^{-1}(l_k)$ from $\Lambda_j^{-1}=\tilde{L}\tilde{L}^T=\sum_{k=1}^pl_kl_k^T$.

\subsubsection{Full conditionals of pooled shrinkage parameters}

The unstructured covariances $\boldsymbol{\Psi}$ are shrunk across-groups towards a pooled unstructured covariance.
The full conditional of the pooled covariance is $\Psi_0|\boldsymbol{Y},\boldsymbol{\theta}_{-\Psi_0}\sim
    \text{Wishart}_p\big(\big(
(\nu-p-1)\sum_{j=1}^J\Psi_j^{-1}+\left(P_{2}\kron P_{1}\right)^{-1}\xi\big)^{-1},\xi+J\nu\big).$
The final level of the hierarchy allows for shrinkage of this pooled unstructured covariance towards a pooled Kronecker structured covariance.
The full conditional for the pooled row covariance is
\[ 
P_{1}^{-1} \sim  \text{Wishart}_{p_1}\big(\left(\xi\sum_{k=1}^p L_k P_2^{-1}L_k^T + P_{01} (\eta_3-p_1-1) \right)^{-1},\eta_3+\xi p_2 \big)
\]
for 
$L_k=vec^{-1}(l_k)$ from $\Psi_0^{-1}=\tilde{L}\tilde{L}^T=\sum_{k=1}^pl_kl_k^T$,
and the full conditional for the pooled column covariance is
\[
P_2^{-1}\sim  \text{Wishart}_{p_2}\big(\left(\xi\sum_{k=1}^p L_k^T P_1^{-1}L_k +P_{02}(\eta_4-p_2-1) \right)^{-1},\eta_4+\xi p_1 \big)
\]
for $L_k=vec^{-1}(l_k)$ from $\Psi_0^{-1}=\tilde{L}\tilde{L}^T=\sum_{k=1}^pl_kl_k^T$.
The corresponding degrees of freedom term may be sampled from a Metropolis step, similar to $\nu,\gamma$. Specifically, a proposal sample may be obtained from a reflecting random walk based on an initial value drawn from
$ \text{Uniform}(\xi-\delta_\xi,\xi+\delta_\xi)$.  Then, $\xi^*$ is accepted as an updated value for $\xi$ with probability
\[
r = \prod_{j=1}^J\frac{p\left(\Psi_0|P_{1},P_{2},\xi=\xi^*\right)p(\xi=\xi^*|r_0,p_0)}{p\left(\Psi_0|P_{1},P_{2},\xi=\xi\right)p(\xi=\xi|r_0,p_0)}.
\]

\subsubsection{A note on computational expense}\label{computationalexpense}

The computational complexity of the proposed Metropolis-Hastings algorithm is at least $\mathcal{O}(\max\{Jp^3,p^2\sum_jn_j,p\sum_jn_j^2\})$.
However, while Bayesian computation of the \mname model may appear cumbersome,  we note that many of the computationally expensive steps may be run in parallel across group. 
Specifically, sampling the random effects $\{U_j\}_{j=1}^J$, the covariance terms $\{\Psi_j\}_{j=1}^J$ and $\{\Sigma_j\}_{j=1}^J$, and the Kronecker covariance terms $\{R_j\}_{j=1}^J$ and $\{C_j\}_{j=1}^J$ may each be comp\-uted in parallel across the $J$ groups.
In this way, the proposed algorithm may scale nicely 
with number of groups, depending on computational resources.

\subsection{Hyperparameter specification}\label{priorspec}

In the absence of meaningful external or prior information, 
weakly informative or non-informative priors on all unknown variables can be
considered. 
On the weight $\lambda$, the prior hyperparameters $\alpha,\beta$ may be selected in a way that weakly encourages 
favoring one of the types of shrinkage, within or across groups,
by setting $\alpha=\beta=1/2$. 

Specifying weakly informative hyperparameters for the degrees of freedom priors requires slightly more scrutiny as the impact 
of $\nu,\gamma,$ and $\xi$ on the Wishart prior distributions
will depend on, among other things, covariance dimension $p$. 
In a Wishart distribution, the degrees of freedom parameter controls the concentration of the distribu\-tion around the prior mean. 
A value that is large relative to the covariance dimension $p$ can correspond to considerable concentration, and a value near the lower bound $p+2$ corresponds to limited concentration.
As such, 
we suggest using a weakly informative prior for a degree of freedom by specifying hyperparameters $r_0,p_0$ that allow for nontrivial prior mass on a range from $p+2$ to values large relative to $p$.
In practice, we found it worked well to 
set hyperparameters such that
a large majority of the prior mass is placed on values in the range $[p+2, 2p]$. 
The size parameter $r_0$ may be set such that the degrees of freedom prior mean is a value near the first quantile of this range. 
The prior success probability may be set such that there is a fair amount of dispersion around the mean. 
In analyses with moderate dimension, we use $p_0=0.2$ which corresponds to a degrees of freedom prior variance of 5 times the prior mean. 
In analyses with large dimension, we use $p_0=0.01$ to allow for a larger prior variance.
In both cases, such a prior tends to be right skewed,
whereby more prior mass is placed on small to moderate values in the parameter space while still incorporating nonnegligible prior mass on moderate to large values.

The hyperparameters on the covariance parameters, $\boldsymbol{R},\boldsymbol{C} ,P_1$, and $P_2$, require special attention due to the over-parameterization of the proposed model
and the scale ambiguity property of the Kronecker product. That is, for a scalar $c$ and matrices $A,B$, $(cA\kron B) = (A\kron cB)$.
To deal with these potential ambiguities,
we propose standardizing the data 
 in a pre-processing step before estimating model parameters and 
setting the scale hyperparameters such that the prior mean of these parameters is
the identity matrix. 
The implications of 
this hyperparameter
choice result in an a priori
homoscedastic 
marginal expectation of the within-group covariances, $E\left[\bs_j\right] = I_p$ for each  $j\in\{1,...,J\}$.
The corresponding degrees of freedom hyperparameters are taken as
$\eta_1=\eta_3=p_1+2$ and
$\eta_2=\eta_4=p_2+2$
to 
represent diffuse distributions that maintain finite first moments.
For example, a Wishart prior of this type,  $ \text{Wishart}_{p_1}(I_{p_1}/(p_1+2),p_1+2)$, on $R_1$ 
corresponds to a diffuse distribution with prior
expected value $I_{p_1}$.
Furthermore, this choice of prior suggests weak shrinkage to an isotropic covariance matrix at the lowest level of the hierarchy. 
Weakly shrinking to such a matrix
has motivations in ridge regression
and the regularization discriminant analysis literature, as in \cite{Friedman1989}.

\section{Simulation Studies}\label{numerical}

We demonstrate
the performance of the proposed \mname model
by comparing 
the 
accuracy of various covariance estimators obtained under four different 
population covariance regimes.
In particular, each regime features either homogeneous (Ho) or heterogeneous (He) covariances across groups, 
and the covariances in each regime are either Kronecker structured (K) or not Kronecker structured (N).
Our goal in this section is to generally explore results when the true group-specific covariance matrices 
have off-diagonal values relatively far from zero
within a given group.
Specifically, 
in unstruct\-ured regimes, each group's true covariance is an exchangeable correlation matrix of dimension $p\times p$ with a fixed correlation randomly generated between 0.35 and 0.9. 
In Kronecker structured regimes, each group's true covariance is the Kronecker product of 
a $p_2\times p_2$ exchangeable correlation matrix and 
a $p_1\times p_1$ exchangeable correlation matrix.
For a given regime and parameter size combination, the true covariance matrices do not vary within the simulation.
Details of the within-group true covariances $\{\bs_1,...,\bs_J\}$ for each regime are contained in Contrast Table \ref{truecovs}.
While we do not necessarily expect the truth in real-world scenarios to be one of these extreme cases, this study provides insight into the behavior of the \mname model.

\begin{table}[H]
\centering
\renewcommand{\arraystretch}{1.3}
\begin{tabular}{ |c||c|c| } 
 \hline
  & Ho &He \\ 
  \hline \hline 
K & $\bs_1=\cdots=\bs_J=Z^{(p_2)}\kron Z^{(p_1)}$ & $\bs_1=Z_1^{(p_2)}\kron Z_1^{(p_1)},\cdots,\bs_J=Z_J^{(p_2)}\kron Z_J^{(p_1)}$ \\ 
 \hline 
N & $\bs_1=\cdots=\bs_J=Z^{(p)}$ & $\bs_1=Z_1^{(p)},\cdots,\bs_J=Z_J^{(p)}$ \\ 
 \hline 
\end{tabular}
\caption{
Population covariance assumptions for each of the four regimes considered.
Notionally, $Z^{(p)}$ is an exchangeable correlation matrix of dimension $p\times p$ where the correlation is a fixed value in $[.35,.9]$.}\label{truecovs}
\end{table}

In a simulation study under each regime,
we compare the loss of various covariance estimators under a range of dimensionality and number of groups.
Specifically, we consider $p_1\in \{2,4,8\}$, $p_2 = 3$, and $J \in \{4,10\}$. 
As we are particularly motivated by the small sample size case, we consider within-group sample sizes $n_1=...=n_J=p+1$. 
For each regime, we take the MLEs of the covariances under the simplest correctly specified model as the oracle estimator. 
In total, this includes
the standard sample MLE for each group's covariance $\boldsymbol{S}=\{{S}_1,...,{S}_J\}$, the pooled sample MLE $\hat{S}_{p}$, the separable MLE for each group
$\hat{\boldsymbol{K}}=\{\hat{K}_1,...,\hat{K}_J\}$, and the pooled separable MLE $\hat{K}_{p}$.
These oracle estimators for each group's covariance under the four regimes considered are specified in Contrast Table \ref{oracleests}.
We note that 
the most accurate estimator of the He,U regime may vary depending on the problem dimension and the number of the groups due to high variance of the sample covariance estimator when sample size is small relative to number of features, as in this study.

\begin{table}[H]
\centering
\renewcommand{\arraystretch}{1.3}
\begin{tabular}{ |c||c|c| } 
 \hline
  & Ho & He \\ 
  \hline \hline 
 K & $\hat{K}_{p}$ & $\hat{K}_1,...,\hat{K}_J$ \\ 
 \hline
 N & $\hat{S}_{p}$ & ${S}_1,...,{S}_J$\\ 
 \hline
\end{tabular}
\caption{Oracle estimators of the population covariances in the four regimes considered.}\label{oracleests}
\end{table}

The simulation study proceeds as follows.
For each permutation of regime, dimension, and number of groups, 
we sample 50 data sets each from a mean-zero matrix normal distribution of the appropriate dimension with population covariance as given in Table \ref{truecovs}. 
For each data set, we compute the reference covariance estimates $\boldsymbol{S}$, $\hat{S}_{p}$, $\hat{\boldsymbol{K}}$, and $\hat{K}_{p}$.
Additionally, we run the proposed Metropolis-Hastings sampler for the \mname model for 28,000 iterations 
removing the first 3,000 iterations as a burn-in period and 
saving every 10th iteration as a thinning mechanism. 
From the resulting 2,500 Monte Carlo samples, we obtain 
the Bayes estimate under an invariant loss, Stein's loss, of each group's covariance, $\hat{\boldsymbol{\bs}} =\{\hat{\bs}_1,...,\hat{\bs}_J\}$ where $\hat{\bs}_j= E[\bs_j^{-1}|Y_j]^{-1}$.
Then,
we compute Stein's loss averaged across the populations 
$\bar{L}(\boldsymbol{\bs},\hat{\boldsymbol{\bs}}) = \frac{1}{J}\sum_{j=1}^JL_S(\bs_j,\hat{\bs}_j)$ where
$L_S(\bs_j,\hat{\bs}_j) = tr\left({\bs}_j^{-1}\hat{\bs}_j\right)-\log\left|{\bs}_j^{-1}\hat{\bs}_j\right|-p_1p_2$.
We report the average of the 50 $\bar{L}$ values
to approximate frequentist risk for each scenario considered. 
In our analysis of results, we refer to $\bar{L}$ as the loss and do not discuss group-specific Stein losses.

In general, we expect inference with the `oracle' estimator to outperform that of the \mname model in each regime considered.
However, as knowledge of
true structural behavior is rare in practice,
the oracle estimator of one regime
obtained from a correctly specified model
may perform arbitrarily poorly in a different regime.
In contrast,
due to the flexibility of the proposed \mname model, we expect
the \mname estimator
to perform nearly as well as each regime's oracle estimator, and outperform the other estimators considered.
Specifically, the \mname model is correctly specified for all four cases as each regime corresponds to particular limiting choices of parameters in the \mname model.
Furthermore, given that we consider problem dimension size similar to each population's sample size, we expect the \mname estimator
to outperform the sample covariance estimators in all cases.

\begin{table}[H]
\centering
\begin{tabular}{ |p{2.12cm}||x{.63cm}|x{.63cm}|p{.63cm}|x{.63cm}|p{.63cm}||x{.63cm}|x{.63cm}|p{.63cm}|x{.63cm}|p{.63cm}|} 
 \hline &&&&&&&\\[-1em]
 & $\hat{\boldsymbol{\bs}}$  & ${\boldsymbol{S}}$ & $\hat{S}_{p}$&  $\hat{\boldsymbol{K}}$ &  $\hat{K}_{p}$& $\hat{\boldsymbol{\bs}}$  & ${\boldsymbol{S}}$ & $\hat{S}_{p}$&  $\hat{\boldsymbol{K}}$ &  $\hat{K}_{p}$  \\
   \hline \hline
  \end{tabular}
\begin{tabular}{ |p{2.12cm}||p{.63cm}|p{.63cm}|p{.63cm}|p{.63cm}|a||p{.63cm}|p{.63cm}|p{.63cm}|a|p{.63cm}|} 
    \multicolumn{1}{|c||}{} & \multicolumn{5}{|c||}{Homogeneous, Kronecker} & \multicolumn{5}{|c|}{Heterogeneous, Kronecker}\\
 \hline
J = 4, p = 6 & 1.36 & 6.82 & \textbf{0.85} & 1.77 & \textbf{0.31} & \textbf{1.66} & 6.82 & 2.54 & \textbf{1.77} & 1.95 \\ 
  J = 4, p = 12 & 2.40 & 13.42 & 1.69 & \textbf{1.36} & \textbf{0.28} & \textbf{2.73} & 13.42 & 5.33 & \textbf{1.36} & 3.74 \\ 
  J = 4, p = 24 & 3.21 & 25.77 & 3.47 & \textbf{1.81} & \textbf{0.44} & \textbf{4.38} & 25.77 & 10.76 & \textbf{1.81} & 7.42 \\ 
  J = 10, p = 6 & 1.14 & 7.12 & \textbf{0.33} & 1.82 & \textbf{0.12} & \textbf{1.54} & 7.12 & 2.21 & \textbf{1.82} & 1.98 \\ 
  J = 10, p = 12 & 2.32 & 13.48 & \textbf{0.65} & 1.39 & \textbf{0.12} & \textbf{2.67} & 13.48 & 4.98 & \textbf{1.39} & 4.37 \\ 
  J = 10, p = 24 & 2.55 & 25.70 & \textbf{1.27} & 1.81 & \textbf{0.17} & \textbf{4.43} & 25.70 & 10.39 & \textbf{1.81} & 9.14 \\ 
 \hline\hline
 \end{tabular}
 \begin{tabular}{ |p{2.12cm}||p{.63cm}|p{.63cm}|a|p{.63cm}|p{.63cm}||p{.63cm}|a|p{.63cm}|p{.63cm}|p{.63cm}|} 
     \multicolumn{1}{|c||}{} & \multicolumn{5}{|c||}{Homogeneous, not Kronecker} & \multicolumn{5}{|c|}{Heterogeneous, not Kronecker}\\
 \hline
  J = 4, p = 6& \textbf{1.49} & 6.82 & \textbf{0.85} & 4.81 & 2.42 & \textbf{1.36} & 6.82 & \textbf{1.96} & 6.27 & 4.53 \\ 
  J = 4, p = 12& \textbf{2.65} & 13.42 & \textbf{1.69} & 7.07 & 5.79 & \textbf{2.77} & 13.42 & \textbf{4.21} & 9.04 & 10.58 \\ 
  J = 4, p = 24& \textbf{4.63} & 25.77 & \textbf{3.47} & 22.02 & 20.27 & \textbf{5.03} & 25.77 & \textbf{8.63} & 31.70 & 31.47 \\ 
  J = 10, p = 6& \textbf{1.47} & 7.12 & \textbf{0.33} & 4.64 & 2.12 & \textbf{1.54} & 7.12 & \textbf{1.48} & 6.14 & 4.44 \\ 
  J = 10, p = 12& \textbf{2.51} & 13.48 & \textbf{0.65} & 7.20 & 5.60 & \textbf{2.63} & 13.48 & \textbf{3.14} & 9.66 & 10.97 \\ 
  J = 10, p = 24& \textbf{3.45} & 25.70 & \textbf{1.27} & 22.32 & 20.27 & \textbf{4.52} & 25.70 & \textbf{6.48} & 33.55 & 33.17 \\ 
  \hline
     \end{tabular}
     \caption{$\bar{L}$ values averaged over 50 iterations for $J$ populations and problem dimension $p=p_1p_2$. The oracle estimator for each regime has a grey background. For each case, the two smallest average losses are in bold font.}\label{simstudy}
 \end{table}

\begin{figure}[htb]
\centering
\includegraphics[width=5.5cm,keepaspectratio]{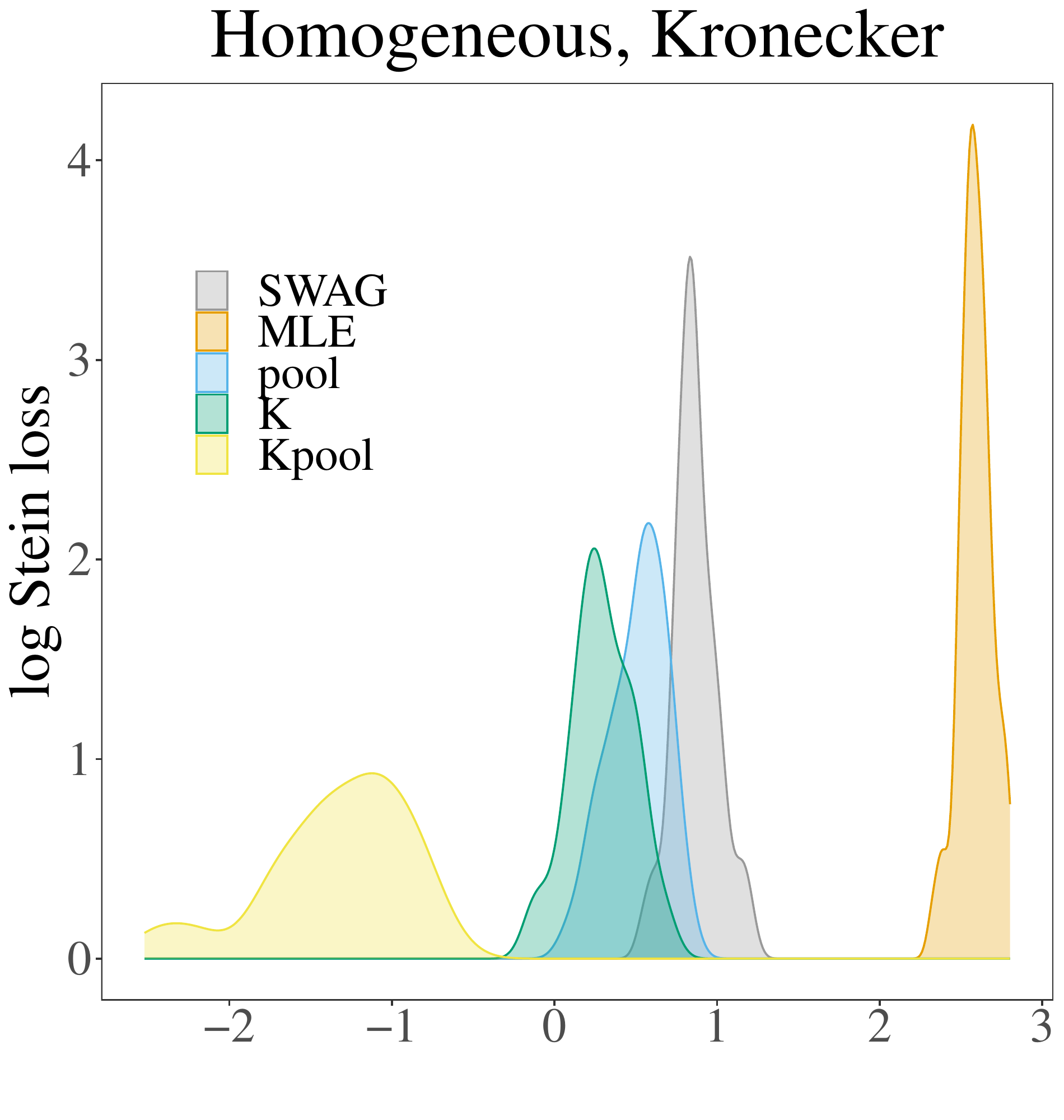}
\includegraphics[width=5.5cm,keepaspectratio]{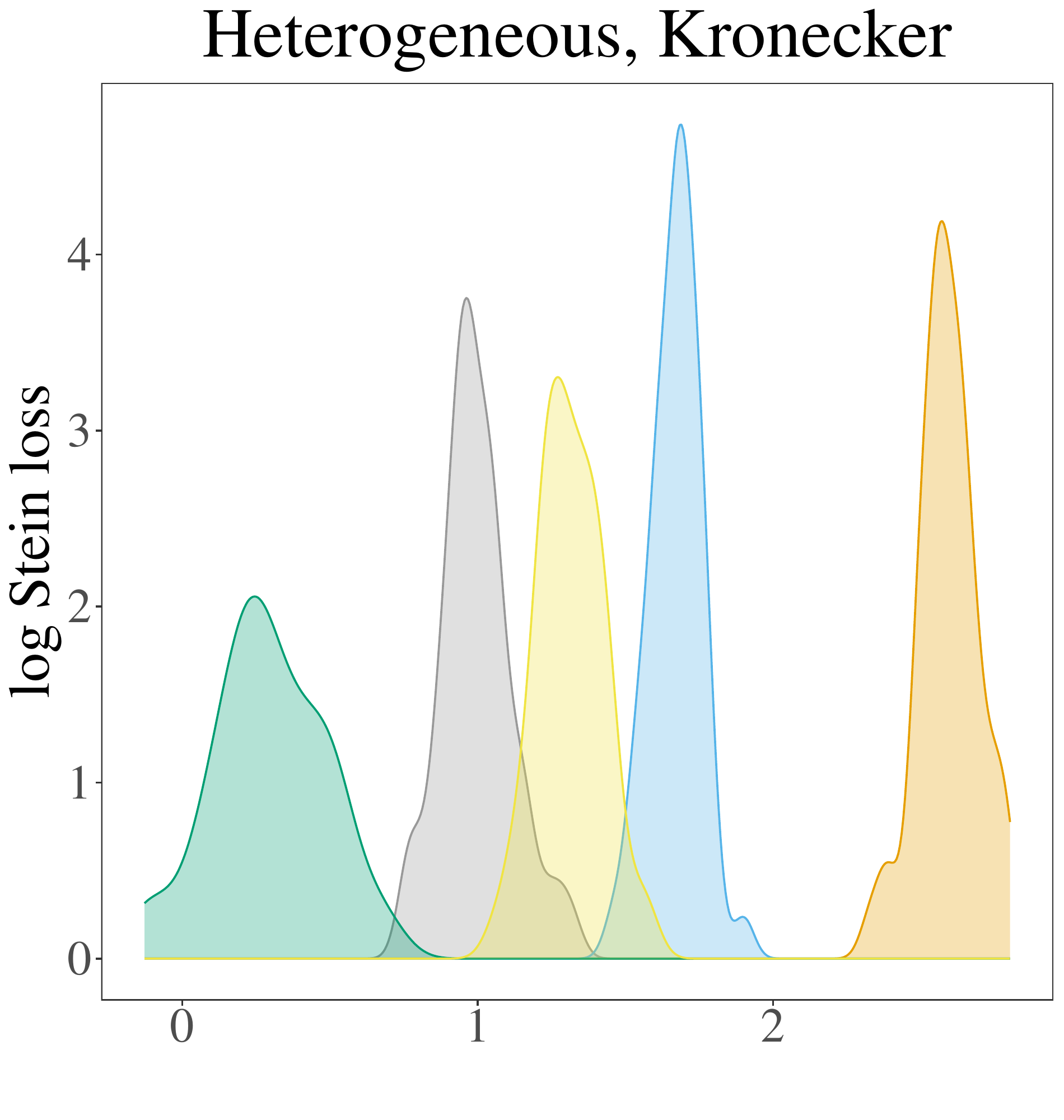}
\includegraphics[width=5.5cm,keepaspectratio]{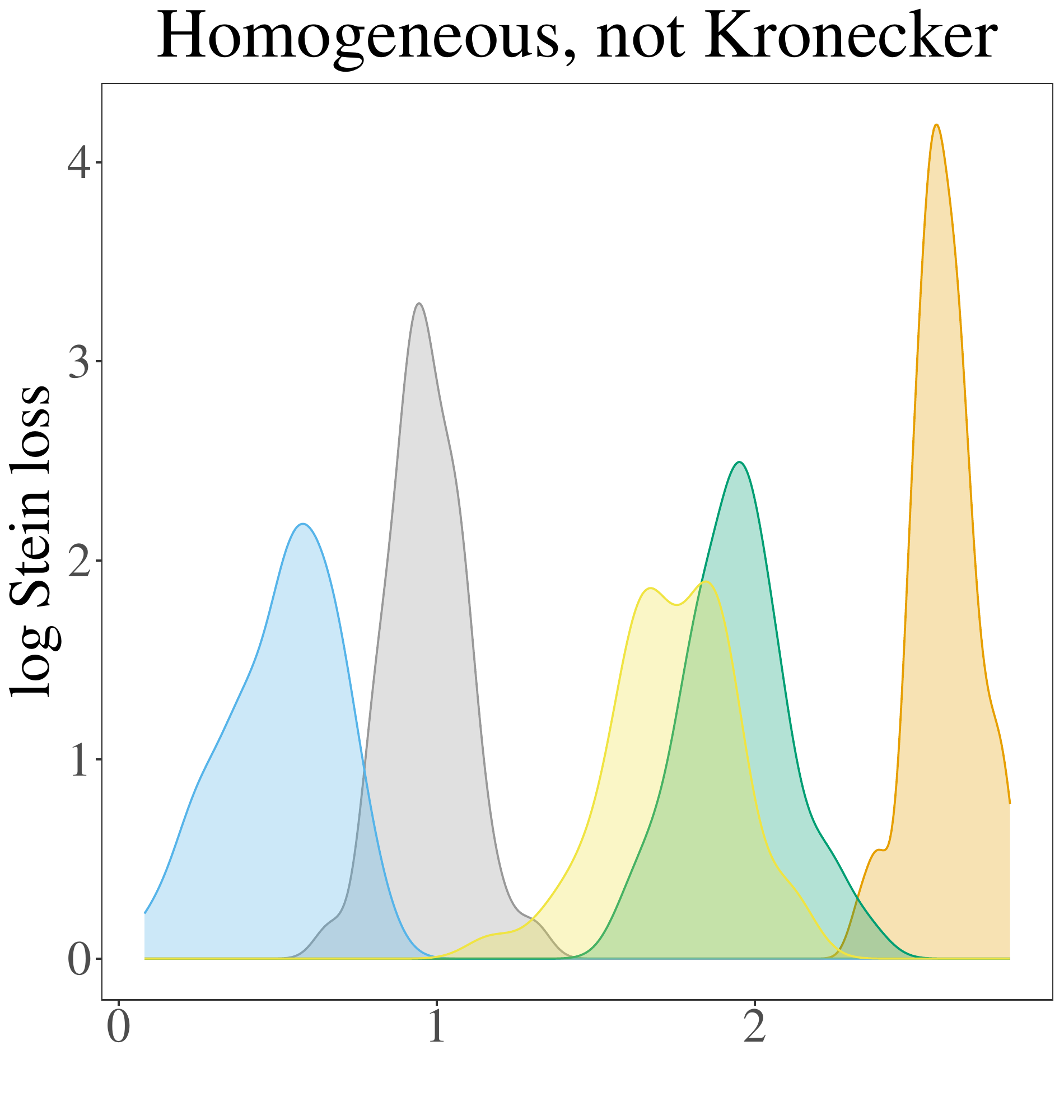}
\includegraphics[width=5.5cm,keepaspectratio]{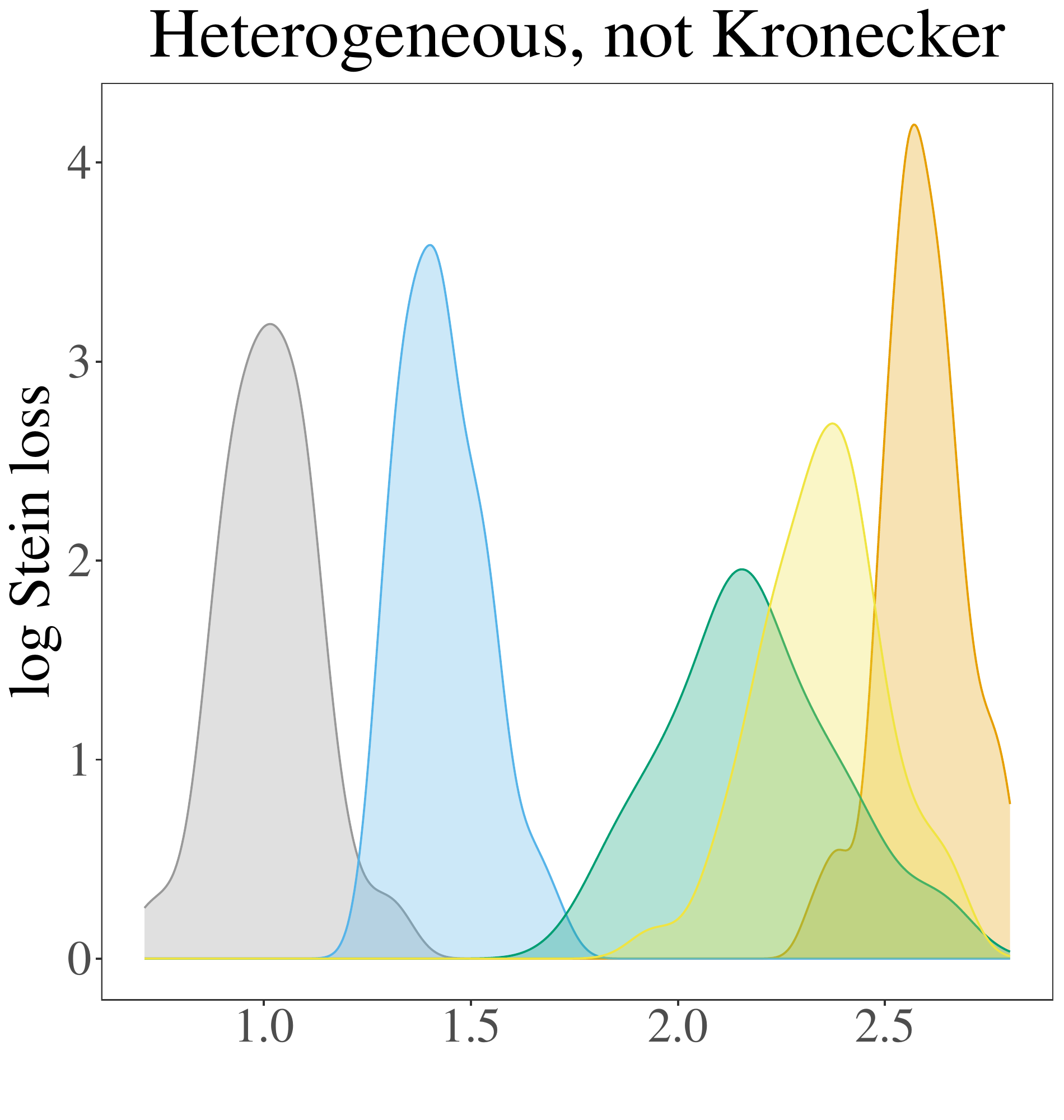}
\caption{
Empirical densities of the 50 $\log(\bar{L})$ values for each estimator given $J=4,p=12$.
}\label{simulation_loss}
\end{figure}

The results of the simulation study are presented in Table \ref{simstudy}. 
In the table, the oracle estimator for each regime has a grey background, and the smallest two average losses are in bold font. In summary, the \mname estimator performs best or second best
in all cases considered except those in the Ho, Kr regime.
In nearly all cases where the \mname estimator has the second smallest loss, it is beat by the respective regime's oracle estimator. 
Therefore, given that the oracle is unknown in practice, we conclude the \mname model is particularly effective in accurate population covariance estimation. 
While the overarching conclusion of the performance of the \mname model remains the same across most cases considered, the dynamics differ across regime.
To gain a better understanding of the variation around the average losses displayed in the table, Figure \ref{simulation_loss} displays the empirical densities of the 50 $\bar{L}$ values for each regime in the case where $J=4,p=12$.

In the regime where population covariances are homogeneous across population and Kronecker structured, the oracle estimator $\hat{K}_{p}$ has the smallest average loss in the cases explored, as expected. 
Interestingly, for a given $J$, the average loss corresponding to $\hat{K}_p$ and $\hat{\mathbf{K}}$ for $p=12$ is less than that for $p=6$. 
This is a consequence of the scaling used within Stein's loss function as,
in this case,
the difference between the first two terms of the loss increases by less than the increase in $p$.
In general, the pooled MLE and group-specific Kronecker MLEs perform well in this regime, 
which is not surprising 
given the variety of estimators considered and the overlap in their underlying assumptions.
The \mname estimator tends to have a similar $\bar{L}$ to these estimators, and a notably smaller average loss than the sample covariance. 
Furthermore, in most cases considered, the empirical density of the loss corresponding to these three estimators ($\hat{\boldsymbol{\bs}},\hat{S}_{p},$ and $\hat{\boldsymbol{K}}$) overlap, as seen in Figure \ref{simulation_loss}.

Given population covariances homogeneous across population and not Kronecker structured, the pooled MLE and \mname estimators have the two smallest average losses.
All other estimators have comparatively large average loss.
This is particularly apparent in Figure \ref{simulation_loss} where the densities of the losses corresponding to $\hat{\boldsymbol{\bs}}$ and $\hat{S}_p$ are smaller than and distinct from the densities corresponding to the other estimators considered.
In high dimension cases in particular, 
incorrectly
assuming a Kronecker covariance in this regime results in estimators nearly as inaccurate as the sample covariances.

In the heterogeneous and Kronecker structured regime, the oracle estimator and the \mname estimator have the two smallest average losses. 
Overall in this regime, these two estimators behave similarly and are notable improvements over the other estimators considered, as seen clearly in Figure \ref{simulation_loss}.
Again, in this regime,
the average loss corresponding to the oracle estimator $\hat{\mathbf{K}}$ for $p=12$ is less than that for $p=6$
as a result of the scaling used in Stein's loss function.

In the fourth, and perhaps most realistic, regime considered, where population covariances are heterogeneous across population and not Kronecker structured, 
the \mname estimator has the smallest average loss in
all cases except when $J=10,p=6$, where it has the second smallest average loss.
In this regime, the oracle estimator is outperformed by all other estimators considered.  
This is not surprising given the 
combination of the simplicity of the exchangeable population covar\-iances in conjunction with large instability of the sample covariance due in part to the small sample sizes considered for each regime.
Again, we see 
estimators based on an incorrect assumption of
separability
result in
distinctly larger loss than the more flexible \mname estimator (Figure \ref{simulation_loss}).

In total, these results support the conclusion that the \mname model outperforms standard alternatives given the true structure of the population covariances is unknown. 
The flexibility of the \mname model is particularly useful given the large error under estimators based on incorrect assumptions. 
In particular, wrongly assuming a Kronecker structure can result in an error nearly as large as that obtained under the sample covariance.
While the extreme regimes considered do not necessarily reflect the truth in real-world scenarios, this simulation study highlights the flexibility of the \mname model.

\subsection{Analysis of computational expense}

\begin{table}[H]
\centering
\begin{tabular}{ |p{2.12cm}||p{1cm}|p{1cm}|p{1cm}|p{1cm}|} 
 \hline &&&&\\[-.75em]
 &Ho,  K   & Ho, N & He, K &  He, N \\
\hline \hline
J = 4; p = 6 & 10.58 & 12.49 & 9.89 & 14.32 \\ 
  J = 4; p = 12 & 13.83 & 16.23 & 14.20 & 18.71 \\ 
  J = 4; p = 24 & 24.46 & 28.36 & 25.07 & 32.59 \\ 
  J = 10; p = 6 & 23.13 & 27.92 & 24.41 & 32.37 \\ 
  J = 10; p = 12 & 31.12 & 36.96 & 31.63 & 42.61 \\ 
  J = 10; p = 24 & 55.92 & 65.04 & 56.37 & 74.33 \\ 
\hline
\end{tabular}
\caption{SWAG Metropolis-Hastings sampler run time in minutes for $S=28,000$ iterations, averaged over 50 replications for each simulation setting comprised of $J$ groups and problem dimension $p=p_1p_2$. }\label{simruntime}
\end{table}

Table \ref{simruntime} displays the wall-clock run time of the Metropolis-Hastings sampler for 28,000 iterations, averaged across 50 replications. 
We sample all parameters as described in Section \ref{fullcond} and do not use any parallelization in the sampling of parameters.  
The algorithm was implemented for a given regime
with 
code written with the 
R statistical programming language 
on the Duke University Compute Cluster on a single thread with 32 CPUs and 228 GB of RAM. 
The smallest parameter space considered ($J=4,p=12$) takes about 10-15 minutes to run, and the largest parameter space considered ($J=10,p=24$) takes about 50-80 minutes.
In summary, doubling of the dimension $p$ from 6 to 12 results in about a 30\% increase in computation time. Doubling $p$ from 12 to 24 results in about a 70\% increase in computation time.
For a given $p$, an increase in the number of groups from 4 to 10 results in an increase in computation time of approximately 125\%.
As discussed in Section \ref{fullcond},  the increase in computation time for large populations may be mitigated by parallelizing sampling across populations in Metropolis-Hastings algorithm.

\section{Examples}\label{examples}

We demonstrate the usefulness of the \mname model 
for estimating covariance matrices in multi-group matrix-variate populations
by analyzing two data sets.
In the first example, we perform a 
speech recognition task 
on a publicly available spoken-word audio dataset.
In the second example, we analyze chemical exposure data
which features small group-specific sample sizes.

In general, a data matrix for a single group $Y$ of dimension $n\times p$ can be decomposed into two orthogonal matrices such that 
one may be used for mean estimation and the other, based on centered data, for covariance estimation. 
As such,
while a mean estimation step could be included in the proposed \mname Metropolis-Hastings algorithm, we will estimate the covariance matrices based on centered data matrices throughout the applications. 
To elaborate,
first, define the $n$-dimensional centering matrix 
$\mathcal{C}=I_n-\mathcal{P}_1$ where $\mathcal{P}_1=\mathbbm{1}_n\mathbbm{1}_n^T/n$ is a rank-1 idempotent projection matrix 
and $\mathcal{C}$ is a rank-$(n-1)$ idempotent projection matrix \citep{Christensen2011}. 
Then, note that,
\[
Y = I_nY-\mathcal{P}_1Y+\mathcal{P}_1Y = \mathcal{C}Y +\mathbbm{1}_n\bar{y}^T
\]
where $\bar{y}$ is the length $p$ vector of column means of $Y$ and $\mathcal{C}Y$ is the residual matrix.
Then, for $Y\sim N_{n\times p}(\mathbbm{1}_n\mu^T,\Sigma \otimes I_n)$, 
$\bar{y}\sim N_p(\mu,\Sigma/n)$ and $nS\sim \text{Wishart}_p(\Sigma,n-1)$ where $S=(\mathcal{C}Y)^T\mathcal{C}Y/n$ is the sample covariance matrix.
$\bar{y}$ and $\mathcal{C}Y$ are uncorrelated, so $\bar{y}$ and $S$ are uncorrelated \citep{Mardia1979}.
In this way, $\Sigma$ may be estimated using centered data $\mathcal{C}Y$.

\subsection{Classification of spoken-word audio data}

Classification of a new observation based on 
a labeled training dataset
consisting of $n_j$ matrices, 
each with common dimensionality $p_1\times p_2$, observed from each of $j\in\{1,...,J\}$ populations is an important statistical task for speech recognition.
To illustrate the utility of the \mname covariance estimator, we 
analyze
classification 
of spoken-audio samples of words “yes”, “no”, “up”, “down”, “left”, “right”, “on”, “off”,“stop”, and “go” from a dataset consisting of 1-second WAV files where
each word has a sample size ranging from 1,987 to 2,103 \citep{Warden2017, Warden2018}.  
Audio data such as these are commonly described by
mel-frequency cepstral coefficients (MFCCs) which represent the power spectrum of a sound across time increments. 
Accordingly, we represent each audio sample as a $p_1\times p_2$ feature matrix of the first $p_1=13$ MFCCs across $p_2=99$ time bins \citep{Ligges2018}.


One popular classification method for generic multivariate data
is quadratic discrim\-inant analysis (QDA).
QDA is based on the result that, assuming normality and equal a priori probabilities of group membership,
the probability of misclassification is minimized by assigning an unlabeled matrix $Y\in\mathbb{R}^{p_1\times p_2}$ to group $j\in\{1,...,J\}$ which minimizes the discriminant score function \citep{Mardia1979},
\begin{equation}\label{qda}
D_j(Y) = (vec(Y)-\mu_j)^T\Sigma_j^{-1}(vec(Y)-\mu_j) + \log|\Sigma_j|,
\end{equation}
where 
$\mu_j\in\mathbb{R}^{p}$ and $\Sigma_j\in\mathcal{S}^{+}_p$, $p=p_1p_2$, are respectively the mean vector and covariance matrix for population $j$
and
$vec(\cdot)$ is the vectorization operator that stacks the columns of a matrix into a column vector. 
In practice, of course, these parameters are unknown and thus are estimated from a training data set.
As a result, adequate performance of the classification relies on, among other things, accurate group-level covariance estimates.

\begin{figure}[H]
\centering
\includegraphics[width=11cm,keepaspectratio]{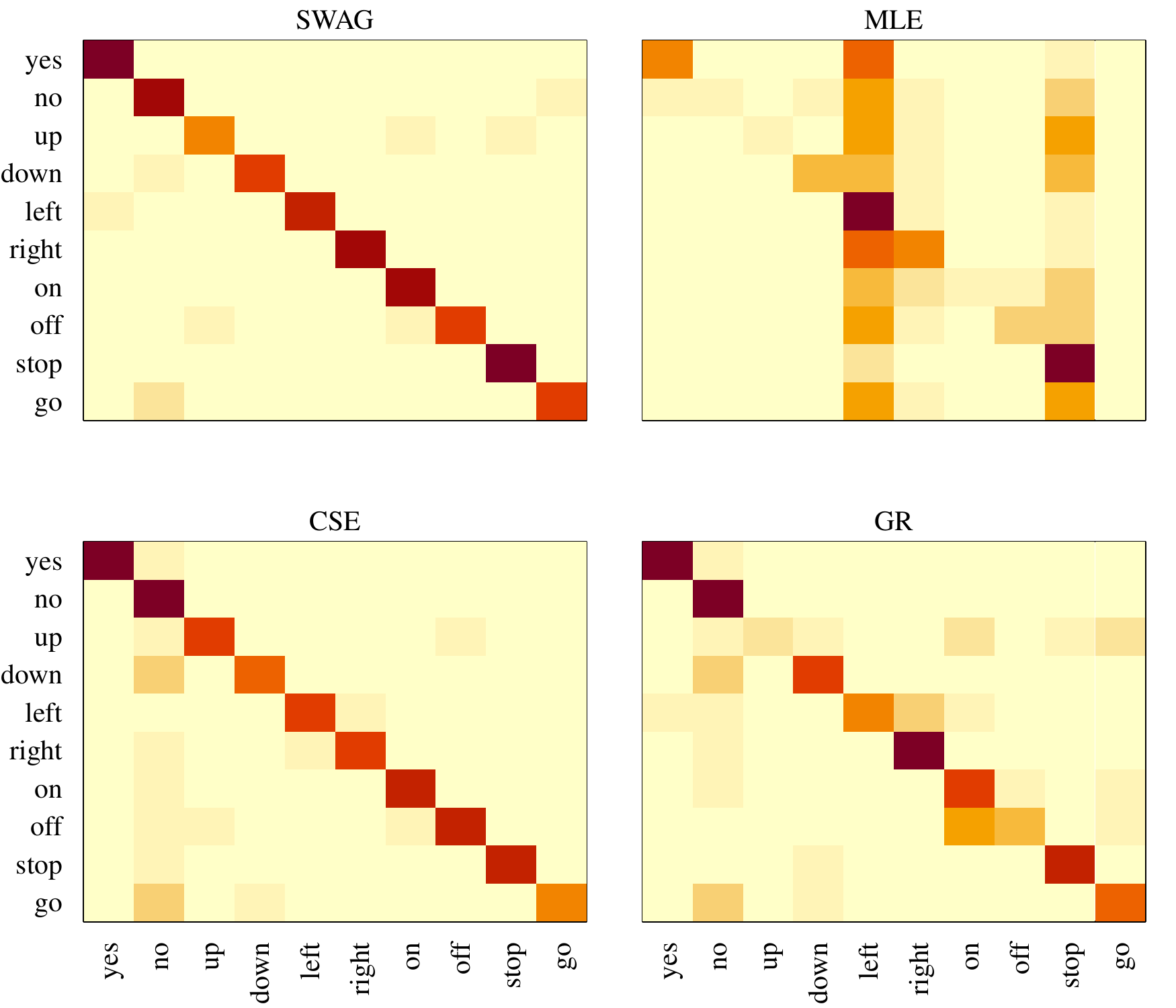}
\caption{Confusion matrices resulting from classification from the various covariance estimates.
Rows correspond to target words and columns correspond to predictions. 
}\label{audioCA}
\end{figure}

We display the utility of the multiple population \mname estimators over 
standard estimators by comparing correct classification rates resulting from QDA in a leave-some-out comparison. 
To do this, we retain a random selection of $100$ observations from each population as a testing dataset, and
the remaining observations constitute the training dataset.
For the discriminant analysis, 
as we are interested in comparing covariance estimation approaches,
we use the sample mean to estimate each $\mu_j$
and a variety of covariance estimates for $\Sigma_j$, all computed from the training dataset. 
Specifically, we obtain the \mname covariance estimates $\hat{\boldsymbol{\bs}}$ from output from the proposed Metropolis-Hastings algorithm run for 5,100 iterations with a burn-in of 300 and a thinning mechanism of 25. 
We compare this with the unstructured sample covariance obtained separately for each word $\boldsymbol{S}$, labeled MLE in this section.
Additionally,
we consider the partially pooled empirical Bayesian covariance estimate outlined in \cite{Greene1989} (GR), and
the core shrinkage estimate \citep{Hoff2022} which partially shrinks each word's sample covariance matrix towards a separable covariance (CSE). 
Prior to computing the various covariance estimates, the data for each word is standardized and centered. The scale is 
then re-introduced to the covariance estimates prior to conducting the classification analysis. 

\begin{figure}[htb]
\centering
\includegraphics[width=4.75cm,height = 4cm]{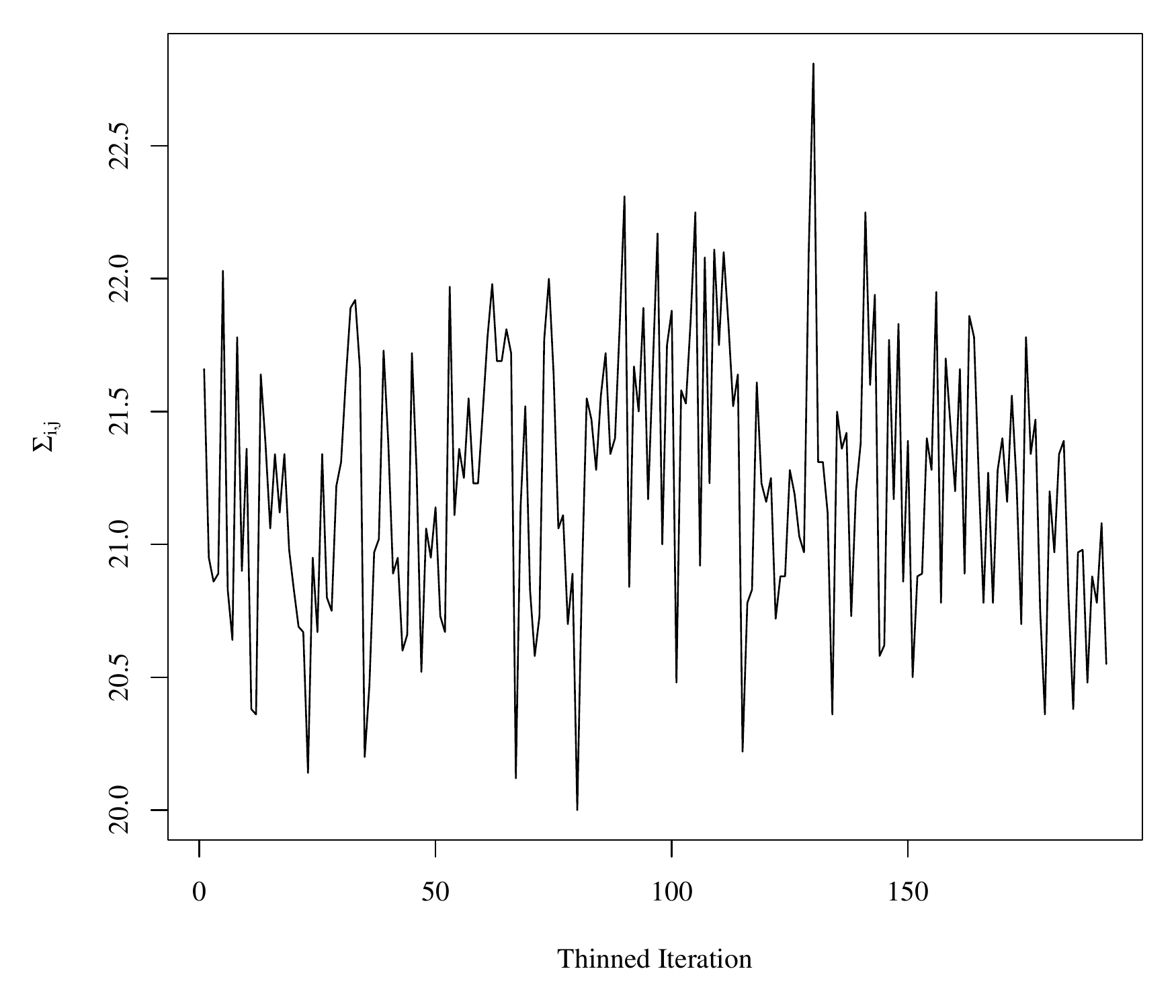}
\includegraphics[width=4.75cm,height = 4cm]{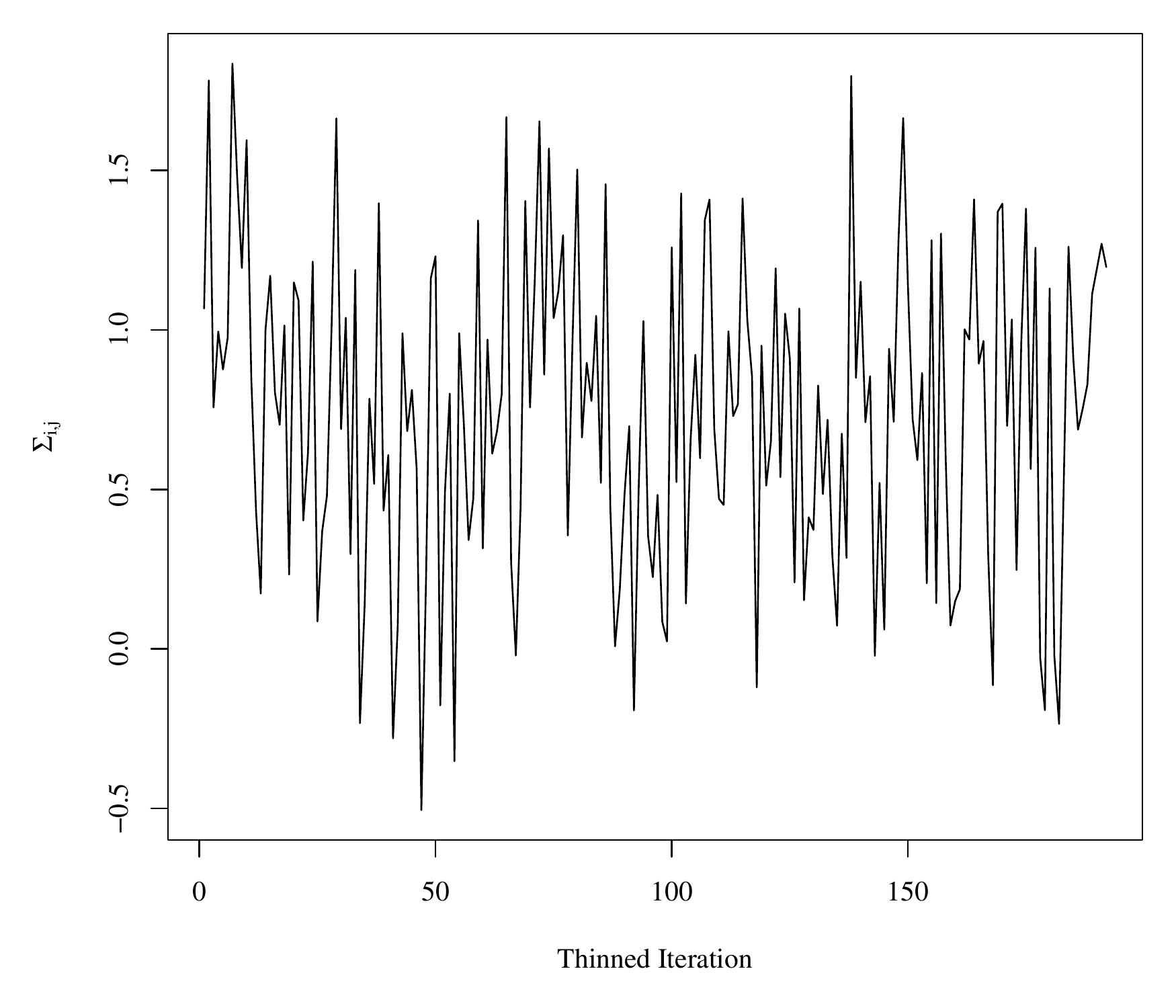}
\includegraphics[width=4.75cm,height = 4cm]{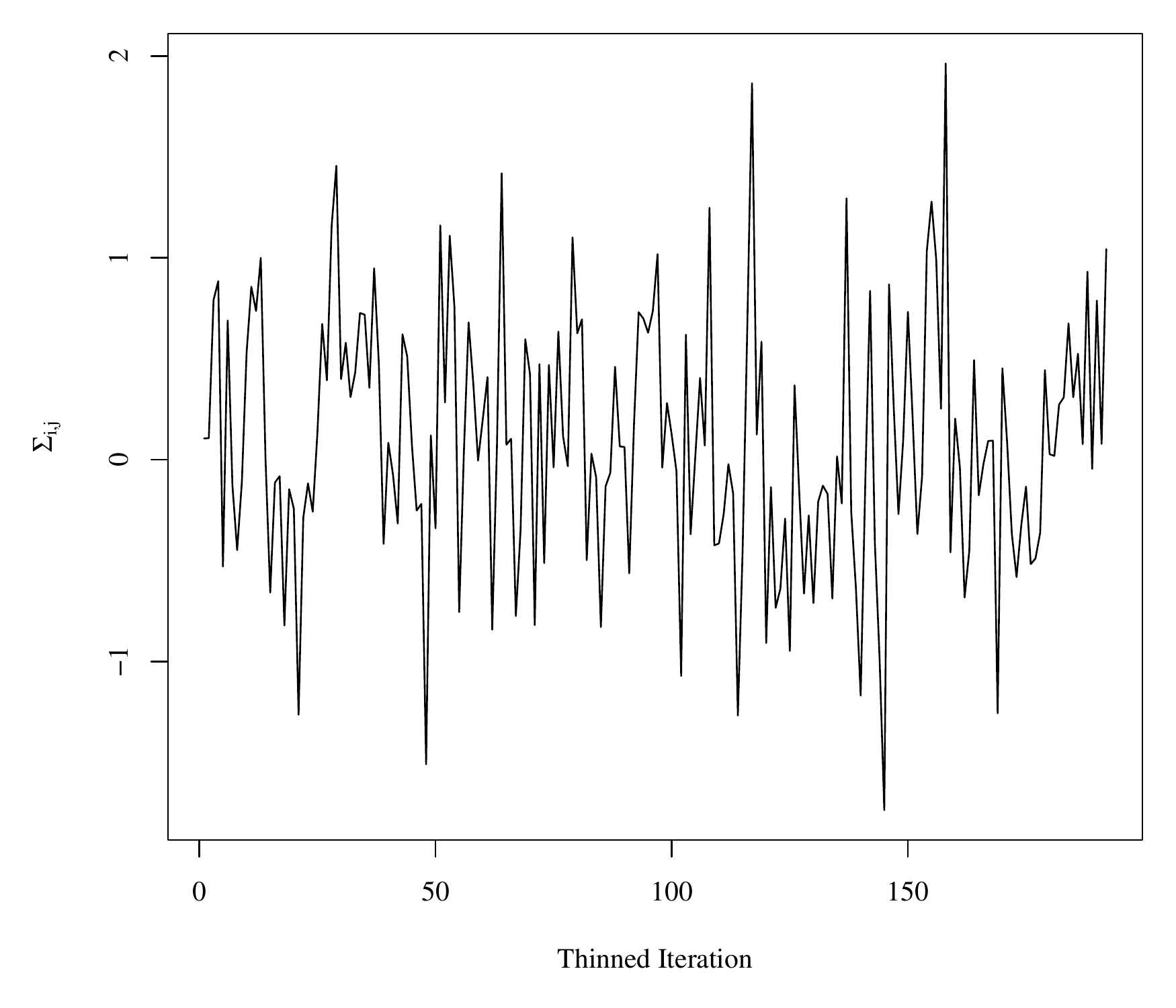}
\includegraphics[width=4.75cm,height = 4cm]{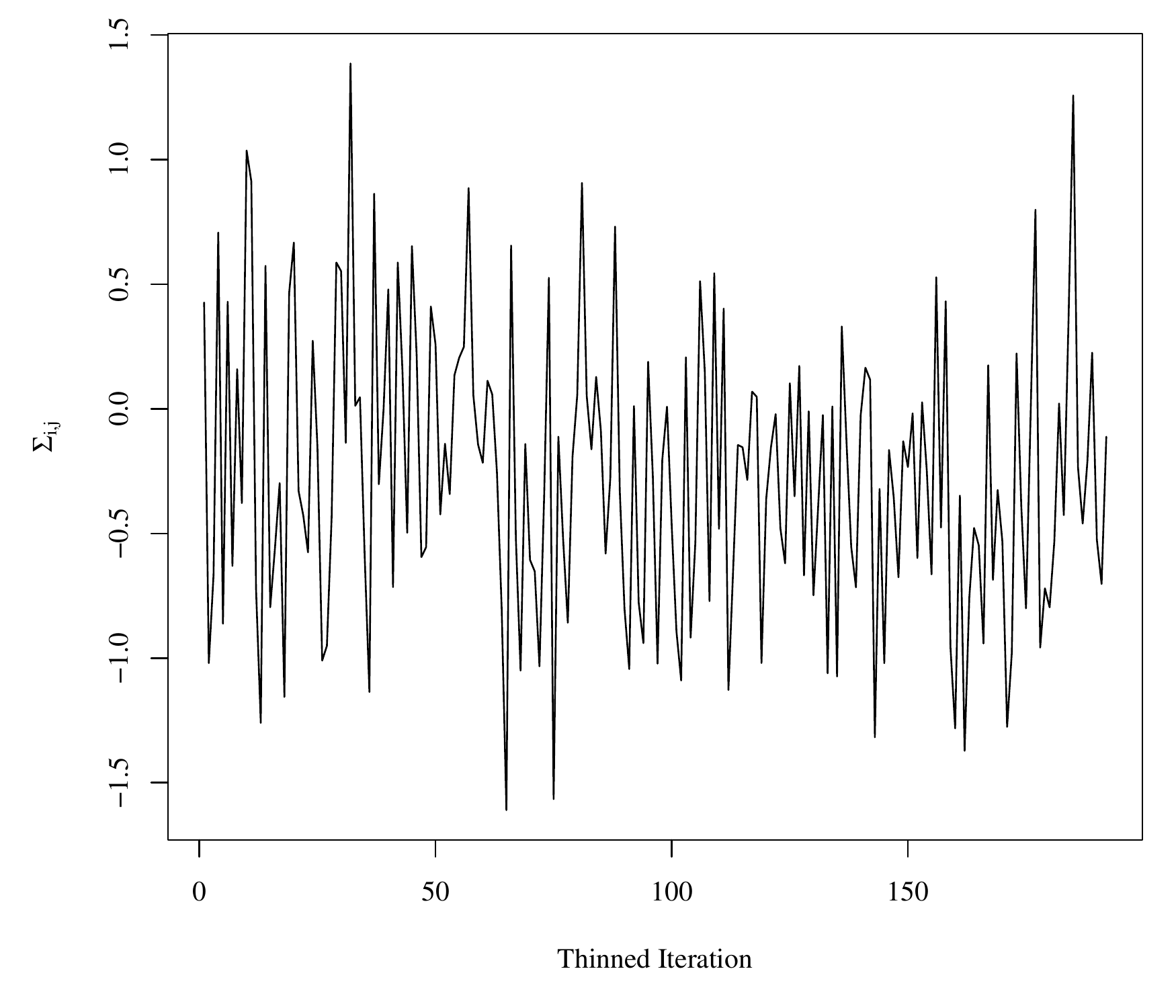}
\caption{MCMC samples for 4 elements selected at random from the set of  covariance matrices $\boldsymbol{\Sigma}$.
}\label{audio_trace}
\end{figure}

To obtain the \mname estimate, we run the Metropolis-Hastings sampler  in an imple\-mentation written in the Julia programming language. 
The code is run on the Duke University Compute Cluster using a single thread with 64 CPUs and 600 GB of RAM. We do not use any parallelization in the sampling of parameters. 
One implementation of the algorithm for 5,100 iterations took approximately 25 hours to run, including compilation of the code and saving the large MCMC output files. 
To assess convergence of the Markov chain, we analyze four randomly selected elements of the within-group covariances $\boldsymbol{\Sigma}$ in detail.
Trace plots corresponding to each of the four elements are plotted in Figure \ref{audio_trace}. 
The maximum lag-10 autocorrelation among the four elements was 0.10 in absolute value,
and the effective sample sizes of the thinned chains corresponding to each element were $117.48,150.72,123.87,$ and $192$ (out of 192 thinned iterations).

\begin{table}[ht]
\centering
\begin{tabular}{ |c||c|c|c|c| } 
 \hline &&&&\\[-1em]
   & SWAG & MLE &CSE & GR \\ 
  \hline \hline
yes & \textbf{0.94} & 0.42 & 0.85 & 0.76 \\ 
  no & \textbf{0.80} & 0.10 & \textbf{0.82} & \textbf{0.80} \\ 
  up & 0.48 & 0.08 & \textbf{0.60} & 0.17 \\ 
  down & \textbf{0.69} & 0.27 & 0.54 & 0.57 \\ 
  left & \textbf{0.74} & \textbf{0.80} & 0.63 & 0.41 \\ 
  right & \textbf{0.86} & 0.43 & 0.58 & 0.77 \\ 
  on & \textbf{0.80} & 0.13 & 0.64 & 0.60 \\ 
  off & \textbf{0.65} & 0.25 & 0.64 & 0.32 \\ 
  stop & \textbf{0.88} & 0.80 & 0.70 & 0.63 \\ 
  go & \textbf{0.64} & 0.05 & 0.47 & 0.47 \\ 
\hline \hline
\textit{average} & 0.75 &  0.33 &  0.65 &  0.55 \\
\hline
\end{tabular}
\caption{Rates of correct classification on audio test dataset from discriminant analysis under different covariance estimators.  The average across all words for each method is displayed in the final row.}\label{audiotable}
\end{table}

Classifications for the test observations are made using each covariance estimate, and results are summarized in confusion matrices displayed in Figure \ref{audioCA}, with the true word classes along the rows and predicted word classes along the columns. 
The correct classification rates, or, the values of the diagonal elements in the confusion matrices, are contained in Table \ref{audiotable}.
In general, the \mname classifier outperforms the other estimates. The \mname estimate has a higher correct classification rate averaged over all words, and it features notably larger word-specific classification rates for the majority of words.

When comparing the \mname estimate with the MLE, it features a significantly higher correct classification rate for every word except two, ``left'' and ``stop''.
Upon further inspection, however, the two large correct classification rates for the MLE are a feature of this classifier nearly always choosing one of these two words, as displayed in the confusion matrix.
The correct classification rates obtained from the \mname classifier are greater than or equal to those from the 
partially pooled estimate GR for every word, oftentimes by a large margin, which corresponds with a greater overall correct classification rate.
Moreover,  linear discriminant analysis, which uses a pooled covariance estimate $\hat{S}_{p}$ for each population covariance, performs even worse with an across-word average correct classification rate of 0.27.
The most convincing competing classifier is the core shrinkage estimate. 
It has a larger correct classification rate for
for the word ``up'' and correctly classifies
two more observations for the word ``no'' than the \mname estimate. While the CSE features slightly better performance for these two words, though, the \mname classifier performs better over all populations. 
Furthermore,  the \mname classifier performs much better overall than the separable MLEs obtained separately for each population,  $\hat{\boldsymbol{K}}$, which has an across-word average correct classification rate of 0.49.
On the whole, the \mname classifier outperforms the other classifiers considered with respect to accurate classification across all populations,
and this example showcases the benefit of allowing for shrinkage both within and across populations.

\subsection{Analysis of TESIE chemical exposures data}\label{tesie_sec}

Many recent medical and environmental studies are concerned with understanding differences among socio-economic groups from repeated measurements of chemical expos\-ures \citep{JamesTodd2017}.
In such an application, researchers may be interested in understanding within and across group heterogeneity. 
Additionally, researchers are often interested in understanding covariate effects and appropriately handling missing data.  
For all such tasks, accurate covariance modeling is critical.

\begin{figure}[htb]
\centering
\includegraphics[width=4.75cm,height = 4cm]{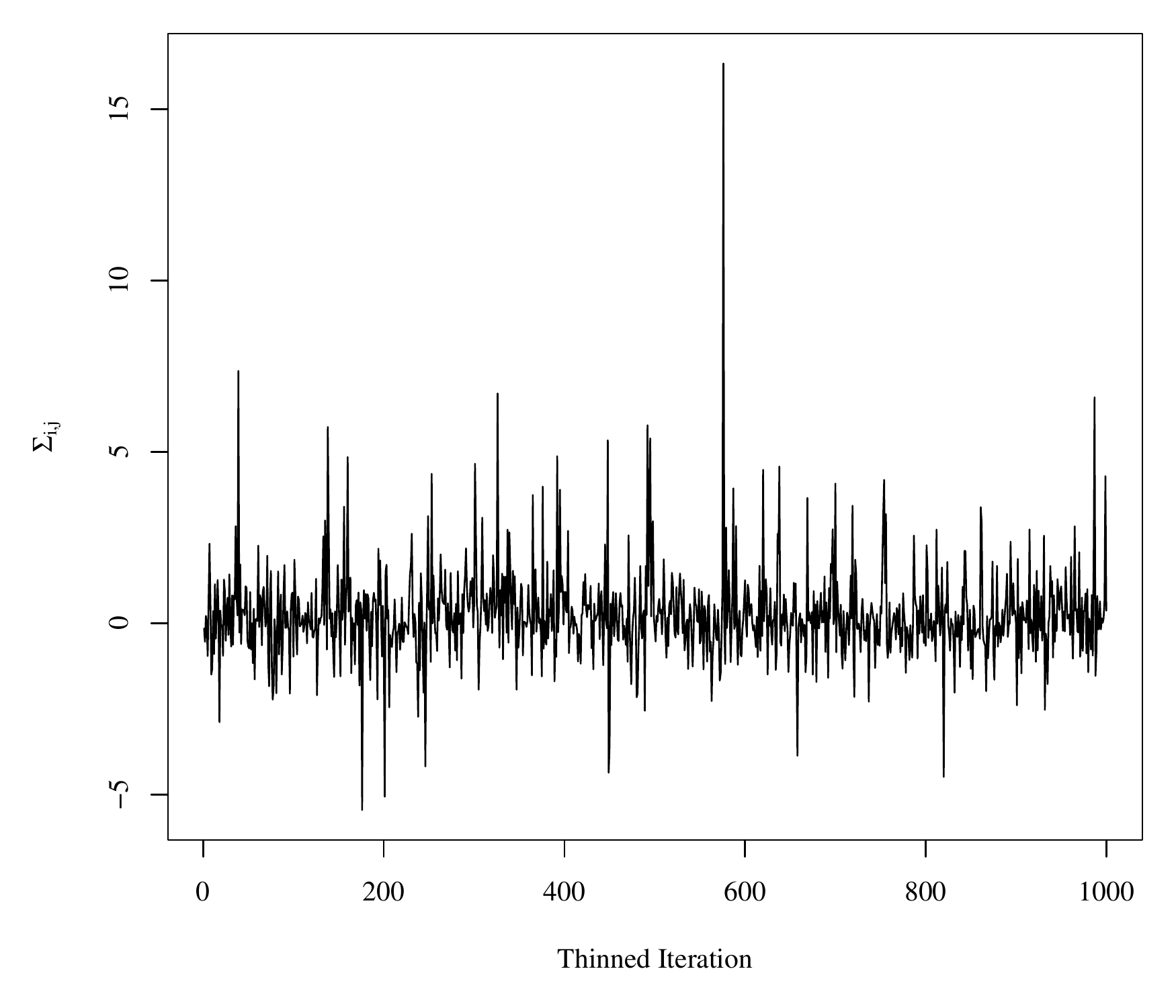}
\includegraphics[width=4.75cm,height = 4cm]{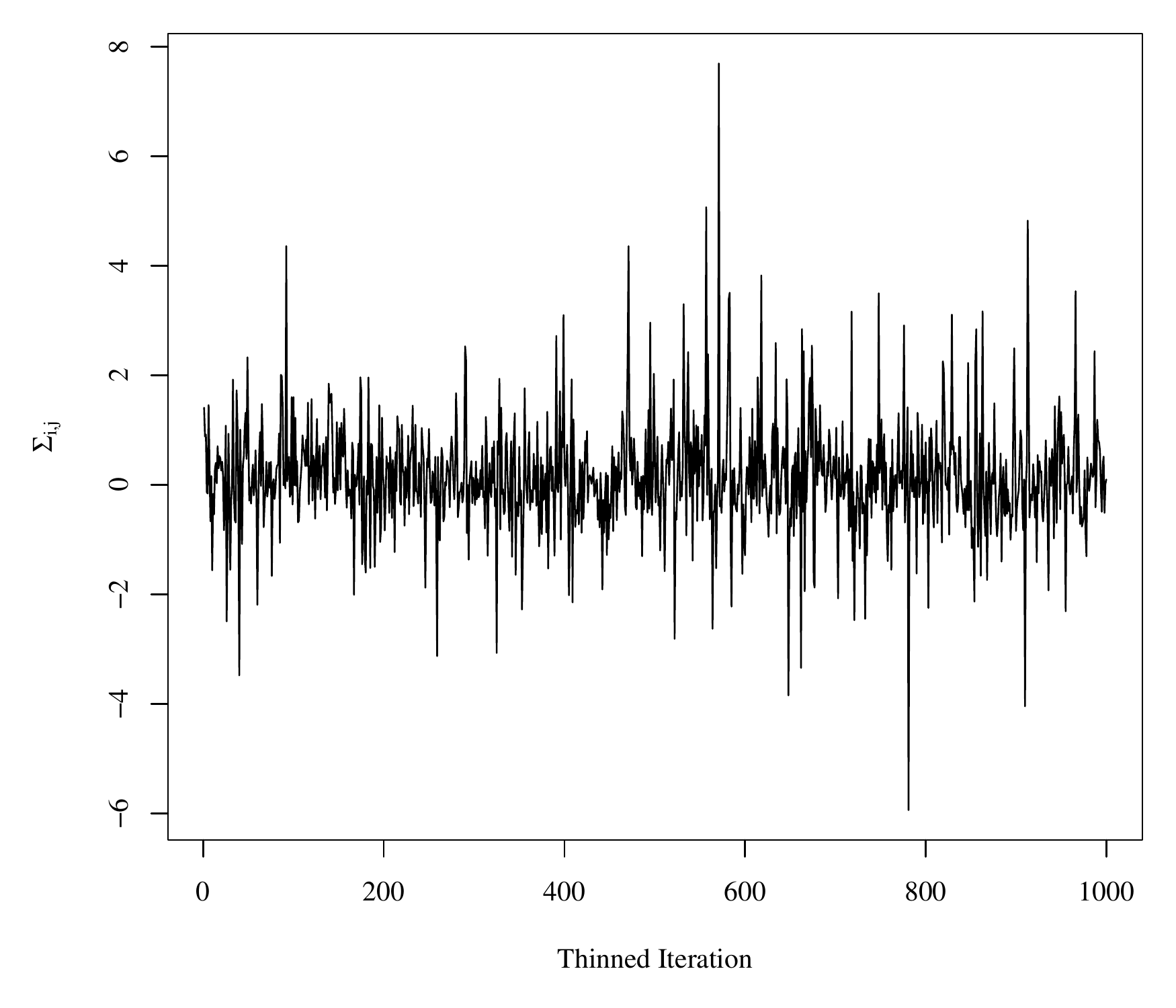}
\includegraphics[width=4.75cm,height = 4cm]{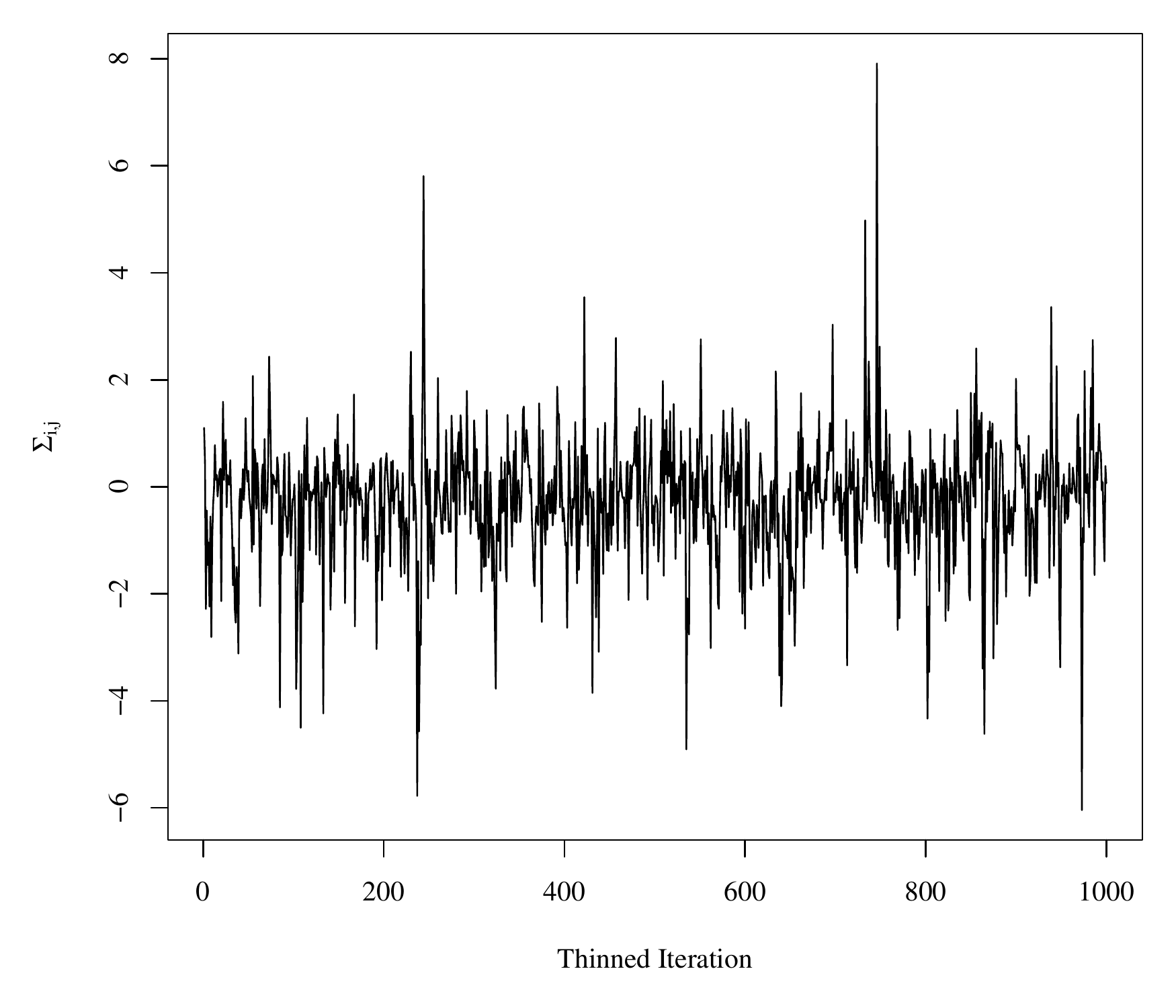}
\includegraphics[width=4.75cm,height = 4cm]{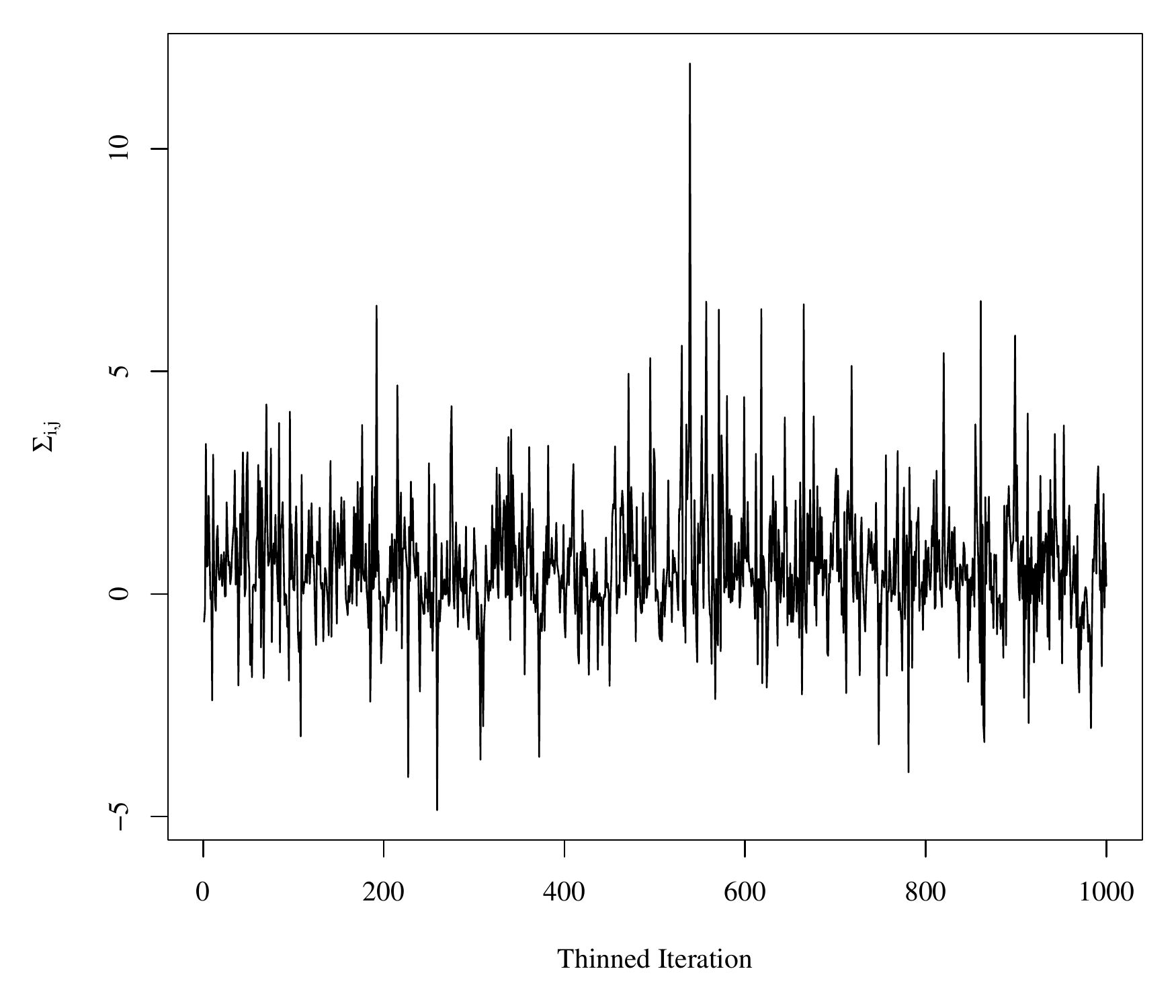}
\caption{MCMC samples for four elements selected at random from the set of  covariance matrices $\{\Sigma_{LHS},\Sigma_{HS}, \Sigma_C\}$.
}\label{tesie_trace}
\end{figure}

In this section, we analyze a sample gathered in the Toddlers Exposure to SVOCS in Indoor Environments (TESIE) study \citep{Hoffman2018}.
In this study, biomarkers of various semi-volatile organic compounds (SVOCs) were extracted from paired samples of urine, blood, and silicone wristbands (see, e.g., \cite{Hammel2018}).
In particular, each observation is a $p_1\times p_2$ matrix where the rows represent biomarkers from $p_1=5$ SVOCs obtained from the $p_2=3$ sources.
Additionally, socio-economic covariates were collected for each individual in the study including highest education level attained and race, among others.
We will analyze the $p_1p_2 \times p_1p_2$ covariance matrices across education level as a proxy for different socio-economic populations. Specifically, the three education levels considered are less than high school (LHS), high school degree or GED (HS), and some college or college degree (C).
The sample sizes are 30, 19, and 24 for the three populations
defined by education levels LHS, HS, and C, 
respectively.

We proceed with simultaneously estimating each group's covariance with the \mname model.
While we remove the mean effect, the \mname model can be extended to include a regression on covariates of interest, for example, 
with the addition of sampling step for a regression coefficient in the proposed MCMC sampler.
We run the Metropolis-Hastings sampler for 33,000 iterations, remove the first 3,000 iterations as a burn-in period, and retain every 30th sample as a thinning mechanism.  
The 33,000 iterations were completed in 3 minutes in an implementation 
of the sampler using the R statistical programming language on a personal machine with an Apple Silicon processor and 8 GB of RAM.
Mixing of the Markov chain for model parameters was good.
The autocorrelation for the thinned chains corresponding to each of the elements in $\boldsymbol{\Sigma}$ was low, with a maximum lag-10 autocorrelation among all 360 elements of 0.11 in absolute value.
Furthermore, the average effective sample size of the thinned chains was 773.73, with a range of 250.40 to 1000 (out of 1000 thinned iterations).  
For reference, see Figure \ref{tesie_trace} for trace plots of four randomly selected elements of $\boldsymbol{\Sigma}$.

\begin{figure}[htb]
\centering
\includegraphics[width=11cm,keepaspectratio]{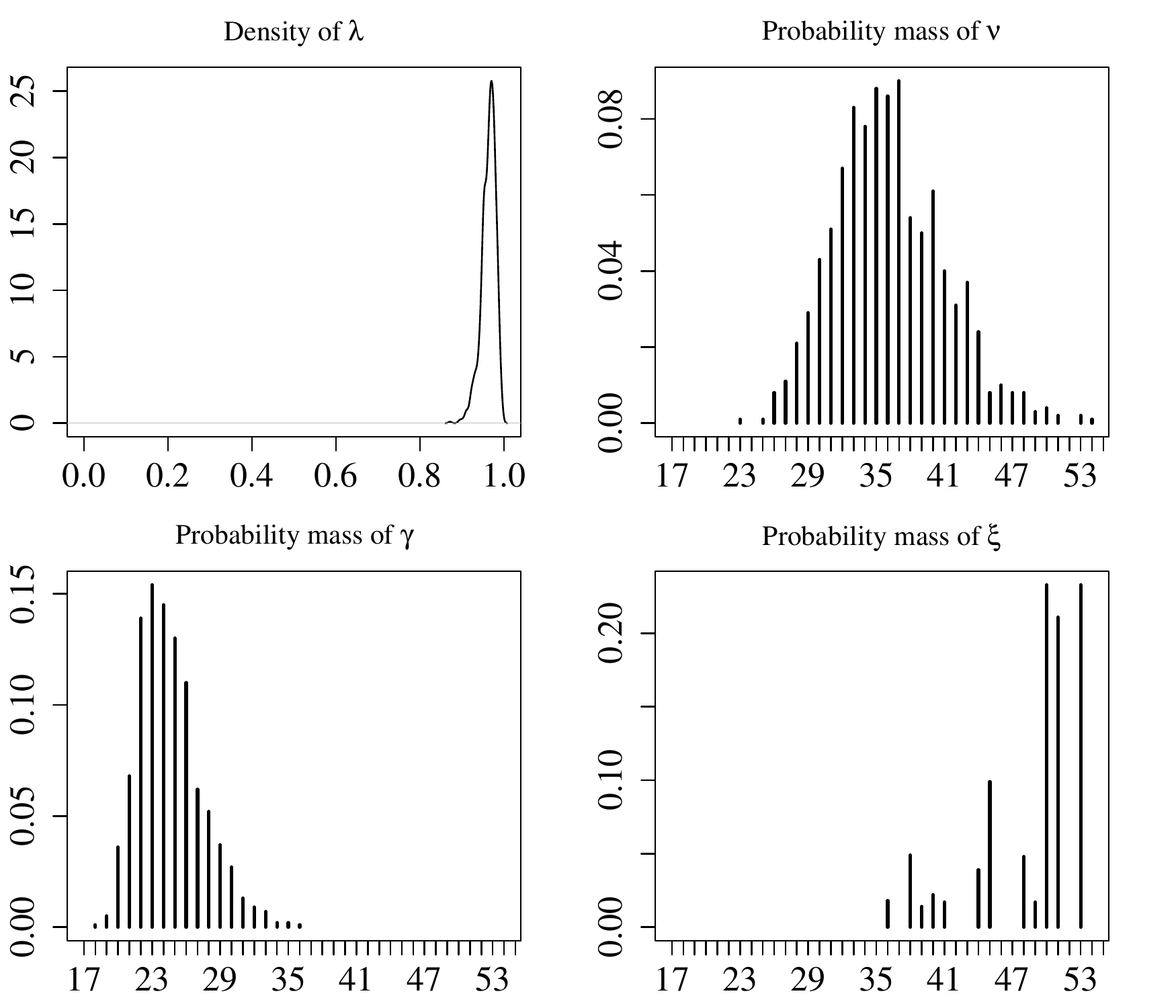}
\caption{Approximations to the posterior distributions of key parameters $\lambda,\nu,\eta,$ and $\xi$ for the TESIE data example. 
}\label{tesie_df}
\end{figure}

\begin{figure}[htb]
\centering
\includegraphics[width=4.75cm,height = 3.95cm]{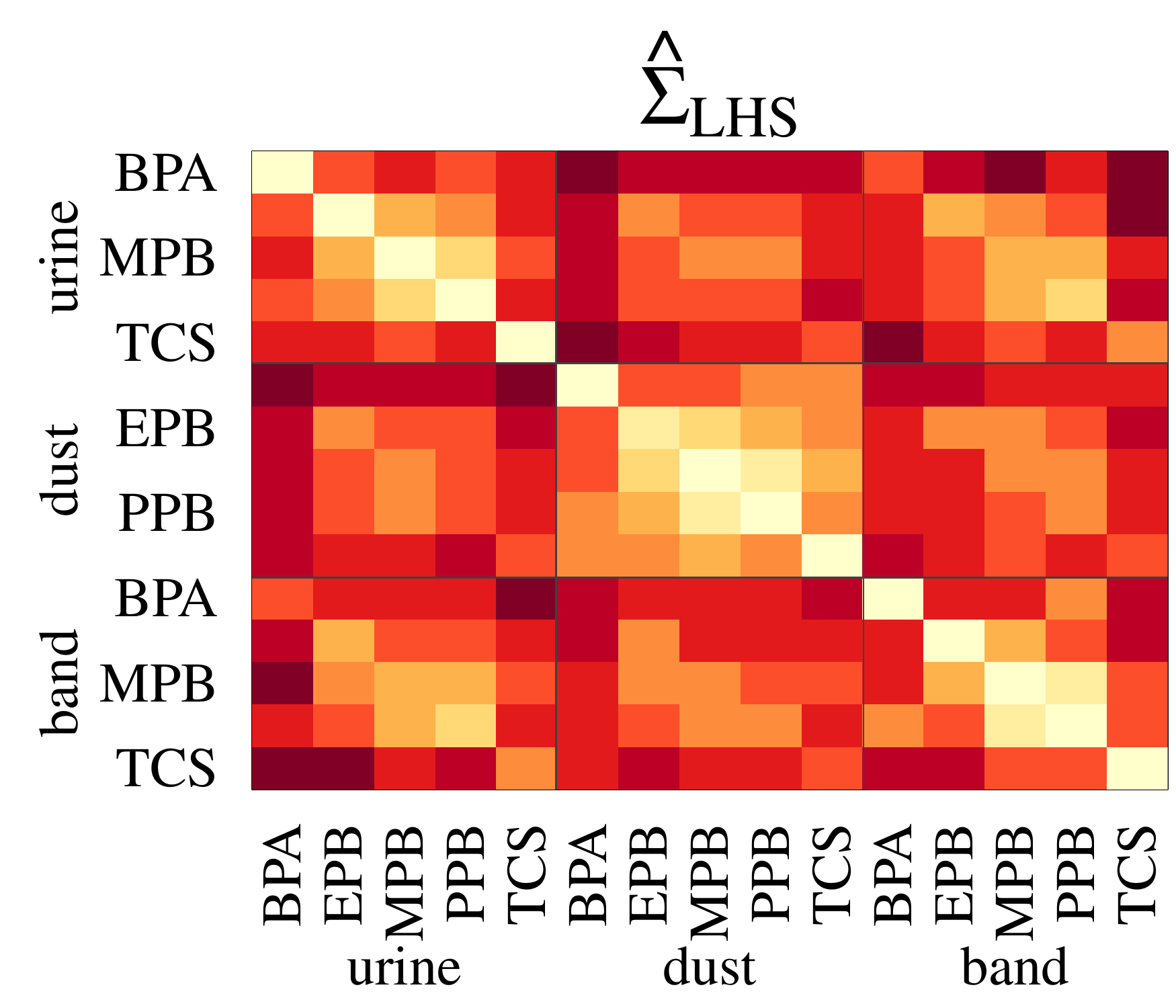}
\includegraphics[width=3.75cm,height = 4cm]{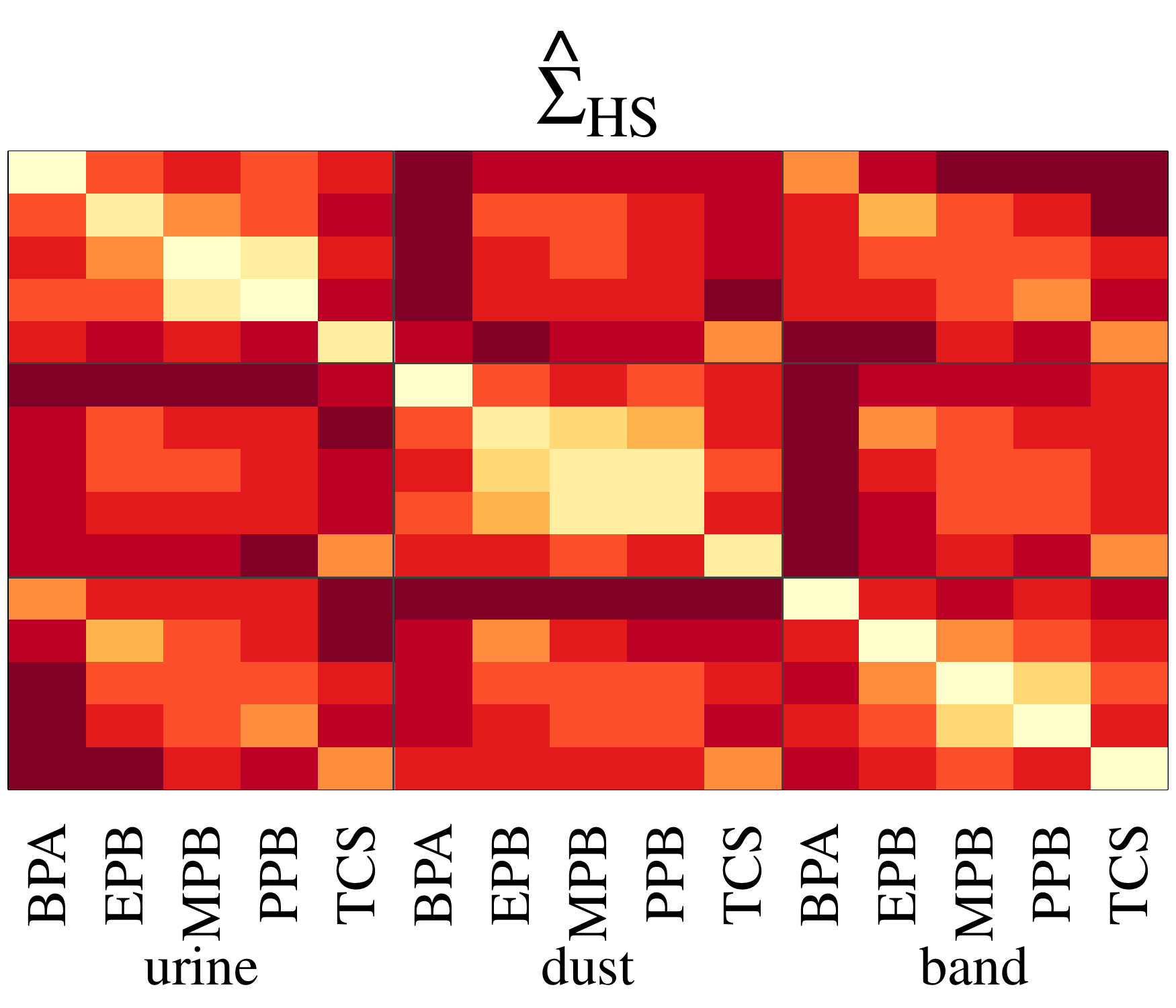}
\includegraphics[width=3.75cm,height = 4cm]{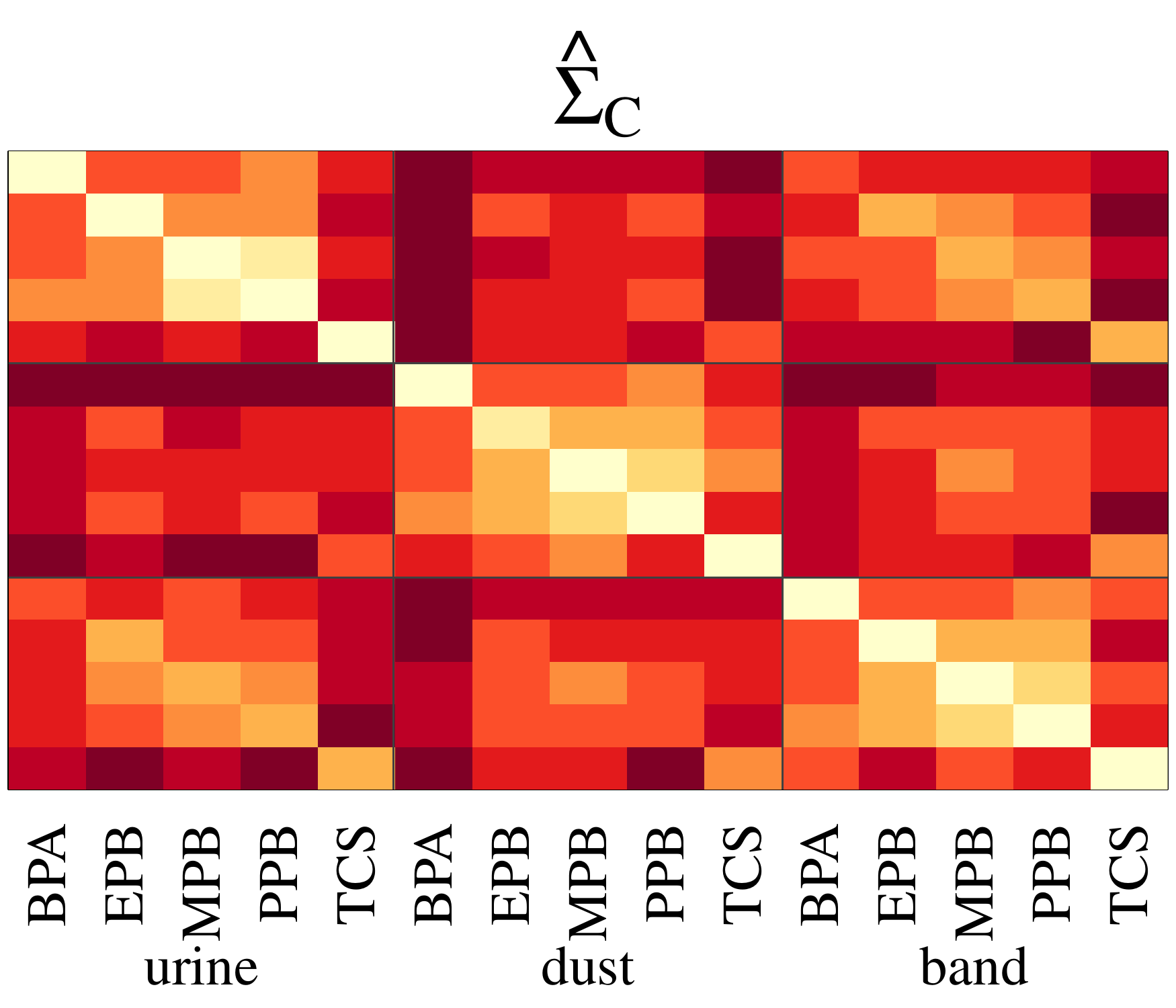}
\includegraphics[width=4cm,height = 4cm]{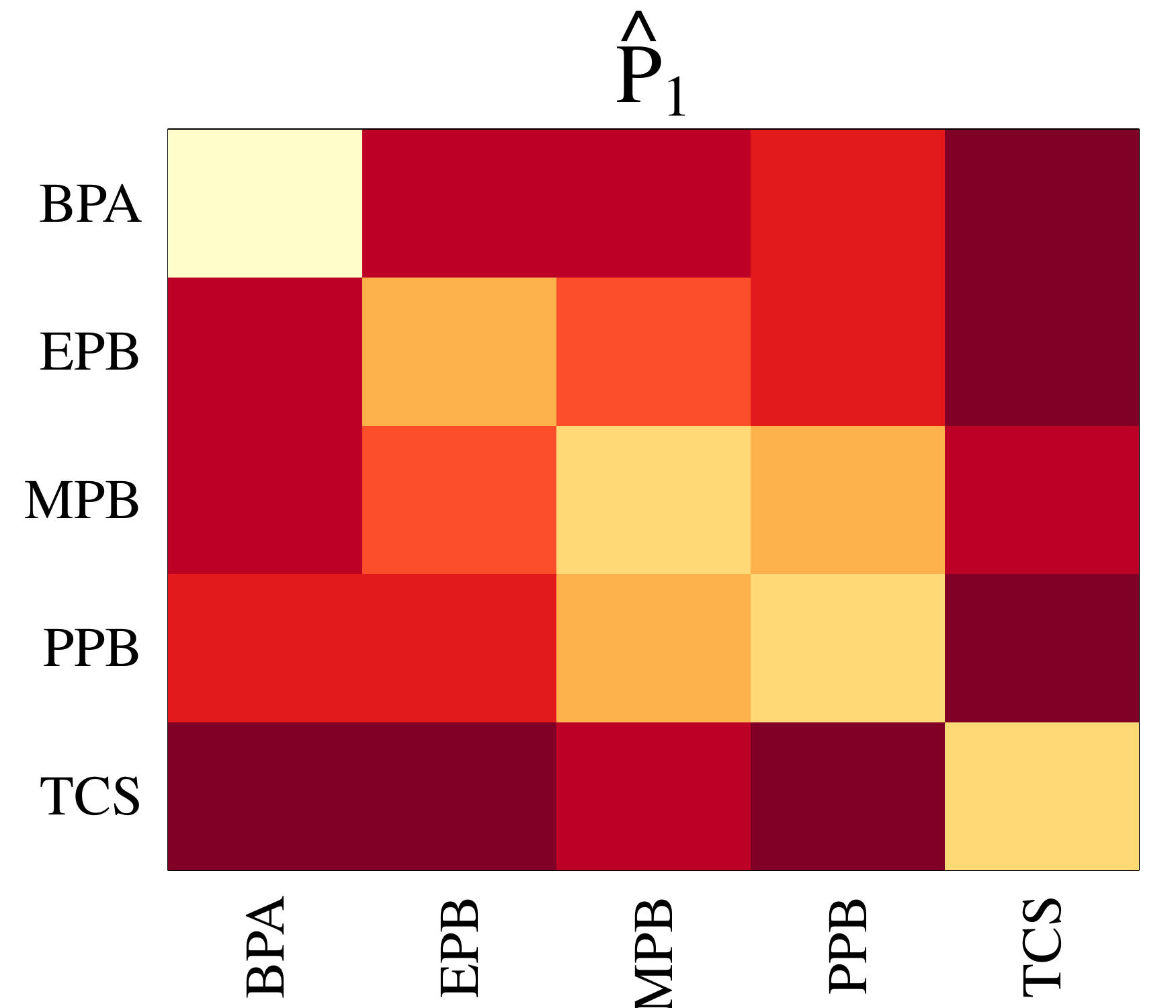}
\includegraphics[width=4cm,height = 4cm]{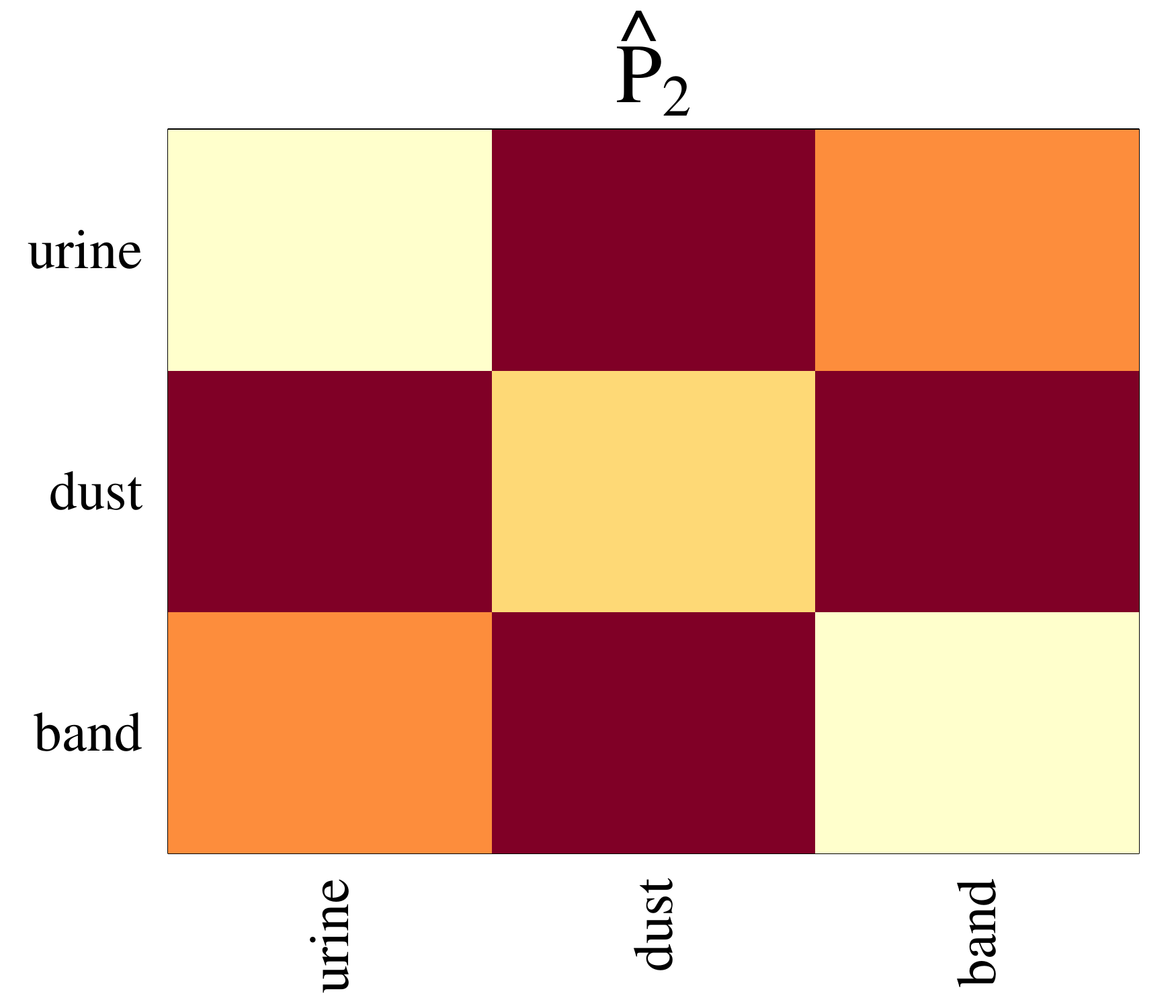}
\caption{Bayes estimates under Stein's loss of the covariances for the three populations LHS, HS, and C are plotted on the top row. Bayes estimates under Stein's loss of the pooled Kronecker covariance are plotted on the bottom row.
}\label{tesie_sigma}
\end{figure}

Approximations to the posteriors of shrinkage-controlling parameters $\lambda,\nu,\gamma,$ and $\xi$ are plotted in Figure \ref{tesie_df}.
The posterior density of the weight on within versus across population shrinkage, $\lambda$, is concentrated around the upper bound of one, indicating that most of the weight is being placed on shrinking across groups towards a pooled covariance. 
Additionally, the posteriors of the degrees of freedom for across-group shrink\-age $\nu$ and $\xi$ are concentrated around large values indicating strong shrinkage from an unstructured covariance towards a pooled Kronecker structured covariance.
In this way, interpretation of summarizing across-row and across-column covariance estimates is straightforward.

This shrinkage behavior is evident upon comparing the Bayes estimates of the group-specific covariances $\hat{\bs}_{LHS},\hat{\bs}_{HS},$ and $\hat{\bs}_{C}$ 
to the Bayes estimates of the pooled Kronecker covariances $\hat{P}_1$ and $\hat{P}_2$ (Figure \ref{tesie_sigma}).
Here, $\hat{P}_1$ is the across-SVOC covariance estimate and $\hat{P}_2$ is the across-source covariance estimate.
The shrinkage towards a pooled separable structure is recognizable throughout the three group-specific covariance estimates. 
In particular, the general pattern of across-SVOC heterogeneity seen in $\hat{P}_1$ is 
roughly present in each 5x5 block in the group-specific covariances, seen in the top row of Figure \ref{tesie_sigma}. 
However, a benefit of the \mname model is the ability to allow for divergences from this separable structure, a feature also discernible in the group-specific estimates. 
For example, within a given population, say, the LHS population,
the pattern among the covariances between BPA and the other SVOCs from samples obtained from urine differs
across measurement source by more than a single factor.
Moreover, this pattern differs 
across populations which reflects heterogeneity across populations.
In total, the output of the \mname model allows for
interpretation of a row covariance and a column covariance, shared across groups, while allowing for deviations from this structure at the group level.

\section{Discussion}\label{discussion}

In this article, we propose a flexible model-based covariance estimation procedure for multi-group matrix-variate data.
The \mname hierarchical model provides a coherent approach to combining two common types of shrinkage, within populations towards a Kronecker structure and across populations towards a pooled covariance.
Bayesian inference of model parameters is straightforward with a Metropolis-Hastings algorithm and allows for uncertainty quantification of covariance estimates. 
In simulation studies, we show
the flexibility of the proposed method results in covariance estimates that outperform standard estimates in a wide array of settings
in terms of loss.

The flexibility of the \mname model is developed specifically for matrix-variate data, but
the model and estimation procedure can be adapted 
for other types of data or applications.
If the data being analyzed are not matrix-variate, different structures can be utilized in place of the Kronecker product.
Additionally, 
the shrinkage towards a pooled covariance can be replaced 
to represent
more complex relationships across the populations such as, for example, an autoregressive relationship.

While an inverse-Wishart prior 
results in computationally convenient inference and 
has a practically useful interpretation 
 in that the
Bayes estimator of the population covariance under a normal-inverse-Wishart hierarchical model is a linear shrinkage
estimator,
it is potentially limiting in that,
for inference on the covariance of a single group, 
one degree of freedom parameter controls concentration around the prior for the $p$ standard deviations and the correlation structure.
As such, a promising direction for future work is to explore SWAG priors on the correlation decomposition of unstructured covariances, as in \cite{Barnard2000}. This could allow for informative priors to be placed on the correlation structure, 
while using uninformative priors on the variable-specific standard deviations, or, vice-a-versa.

Replication codes are available at \url{https://github.com/betsybersson/SWAG}.

\begin{acks}[Acknowledgments]
The authors thank Kate Hoffman and Heather Stapleton for providing data from the Toddlers Exposure to SVOCS in
Indoor Environments (TESIE) study analyzed in Section \ref{tesie_sec}.
\end{acks}

\begin{supplement}
\stitle{Supplement A: Metropolis-Hastings Algorithm for SWAG}
\sdescription{This supplement contains explicit details of a Metropolis-Hastings algorithm for parameter estimation for the SWAG model.}
\end{supplement}

\bibliographystyle{./MiscFiles/ba}
\bibliography{./MiscFiles/library}

\end{document}